\documentclass[review]{elsarticle}

\usepackage{lineno,hyperref}
\usepackage{color}
\usepackage{amsmath,amssymb}
\modulolinenumbers[5]

\journal{Journal of \LaTeX\ Templates}

\usepackage{color}
\usepackage{soul}
\newcommand{\rp}[1]{\textcolor{black}{#1}}

\begin{document}

\begin{frontmatter}

\title{Emergent pseudo time-irreversibility in the classical many-body system of pair interacting particles}

\author{Gyula I. T\'oth}
\address{Interdisciplinary Centre for Mathematical Modelling, and Department of Mathematical Sciences, Loughborough University, LE11 3TU Loughborough, United Kingdom}

\begin{abstract}
\rp{In this paper, the emergence of macroscopic-scale pseudo time-irreversibility is studied in the closed classical many-body system of pair interacting particles. First, exact continuum equations are derived to the Hamiltonian dynamics without the utilisation of statistical mechanics. Next, it is shown that the momentum density field incorporates the thermal degrees of freedom in the considered scaling limit, and therefore initial condition indicated pseudo time-irreversible solutions may exist to the dynamical equations. Numerical evidence for pseudo-irreversible thermal equilibration, heat and momentum transport is provided. Finally, the possible reasons of the numerically obtained non-diffusional relaxation of macroscopic order is discussed.}
\end{abstract}

\begin{keyword}
classical n-body problem, Loschmidt's paradox, time irreversibility, scaling limit
\end{keyword}

\end{frontmatter}


\section{Introduction}

\rp{One of the most perplexing unsolved problems in fundamental physics is the microscopic origin of the thermodynamic arrow of time provided by the second law of thermodynamics.} In particular, while the fundamental microscopic mathematical models of matter only provide time-reversible solutions, \rp{macroscopic-scale} spatial order in a \textit{closed} system is known to solely decay in spontaneous temporal processes, which results in a contradiction known as Loschmidt's paradox \cite{Loschmidt}. {It is obvious that the time-symmetry of the general solution of \rp{the microscopic dynamical equations} is not affected by the number of degrees of freedom, and therefore a closed Hamiltonian many-body \rp{system of pair interacting particles} is time-reversible. Nevertheless, \rp{it is generally accepted that} solutions exhibiting pseudo time-irreversibility may emerge \rp{in a time-reversible system of huge number of degrees of freedom} due to \rp{special initial conditions}. \rp{The explanation of the phenomenon} is \rp{usually} linked to Poincar\'e's recurrence theorem \cite{doi:10.1142/9789812704016_0039}: One can argue that even though the dynamics is recurring, the recurrence time for certain initial conditions \rp{can be much} larger than the estimated age of the Universe, and therefore the time evolution of the system is practically irreversible on the time scales of interest \cite{doi:10.1142/9789812704016_0039,1583121}. \rp{While non-linearities in coupled dynamical equations indeed seem to transfer kinetic energy irreversibly in this sense from macroscopic to thermal scales}, the recurrence theorem \rp{alone} provides no guarantee for the \textit{diffusive} \rp{macroscopic-scale} spatio-temporal relaxation of mass \rp{(chemical diffusion)}, momentum \rp{(viscous flow)} and kinetic energy \rp{(heat conduction)}, the \rp{observable} manifestations of the second law of thermodynamics. 

\rp{Loschmidt's} paradox has been partially resolved in the framework of statistical mechanics. Although the Liouville equation is time-reversible for a time-reversible microscopic dynamics, exact non-equilibrium solutions are unknown in general, which leaves room to \textit{construct} time-symmetry breaking ensembles by \rp{incorporating a Markovian stochastic component}. The most common example is the derivation of the Boltzmann Kinetic Equation, where the Hypothesis of Molecular Chaos \rp{(assuming uncorrelated velocities of colliding particles)} is \rp{utilised} to resolve the two-particle distribution function in the BBGKY hierarchy \cite{Ardourel2017}, \rp{which then leads to the H-theorem}. Another famous example is the Zwanzig-Mori projection formalism, where a Markovian approximation is utilised to resolve a convolution term \cite{Vrugt_2020}. The extension of the Dynamical Density Functional Theory for viscous liquids is also an illustrative example \rp{for imposed time-symmetry breaking} \cite{doi:10.1063/1.4913636}, where the Taylor expansion of the Maxwell-Boltzmann one-particle distribution is used around Local Thermodynamic Equilibrium \cite{kreuzer1983nonequilibrium}\rp{, thus leading to the appearance of a viscous stress tensor in the Navier-Stokes equations}. In a recent approach, an almost exact derivation of the Navier-Stokes equations was provided for initially locally Gibbsian ensembles \cite{PhysRevLett.112.100602}. In this framework, viscosity emerges as the first finite-scale correction to the ideal hydrodynamic limit, \rp{however, the derivation is also based on a Markovian assumption.}

\rp{Apart from the statistical mechanical treatment of Hamiltonian systems of many degrees of freedom,} diffusion can be directly incorporated in dynamical systems by adding a time-continuous stochastic noise to the \rp{otherwise deterministic} equations  \cite{https://doi.org/10.1002/zamm.19820621219,doi:10.1142/9789813279612_0005}. \rp{In addition}, it is known from the analysis of weakly perturbed Hamiltonian systems that incorporating a Markovian stochastic component operating even on a single degree of freedom is \rp{sufficient} to break the time reversibility of the \rp{solution} \cite{Lykov2015,10.1007/978-3-319-65313-6_11,10.1214/18-EJP177}. \rp{A similar technique is utilised} in molecular \rp{dynamics} simulations\cite{DHAR2019393}, since thermostats being utilised to control the temperature in these simulations \cite{Hunenberger2005} makes the otherwise non-ergodic Hamiltonian dynamics \cite{Meyer1974,arnoldbook,Chierchia2011} ergodic \cite{doi:10.1063/1.4792202}, which therefore converges to a statistical mechanical-compatible equilibrium.

Although the results of statistical mechanical based approaches are in excellent agreement with experimental observations, obtaining the desired result may justify the \rp{utilised} assumptions \rp{and approximations} that led to it, but never proves their exactness. \rp{Consequently}, the existence of pseudo time-irreversible solutions as well as the \rp{form} of the relaxation processes in deterministic systems can only be investigated \textit{directly}. Such studies are available for fully chaotic maps, where a statistical phenomenon called ``deterministic diffusion'' has been discovered \cite{Driebe1999}. Regarding the closed Hamiltonian many-body problem, it has already been shown that this system satisfies the Euler equations in the hydrodynamic limit \cite{Olla1993}. Later, a direct derivation of the Euler equations was also given for simple closed Hamiltonian particle systems \cite{Lykov2017}. \rp{These results indicate that the second law of thermodynamics is absent in a closed Hamiltonian system in the exact sense, but still may be present in the the form of initial condition indicated pseudo time-irreversible solutions. To investigate whether these solutions exist on macroscopic scales, here we derive such an exact scaling limit to the closed Hamiltonian many-body system of pair interacting particles, in which the thermal component is preserved by the continuum momentum density, and therefore the effect of special microscopic initial conditions on the macroscopic-scale spatio-temporal behaviour of the system can be studied.}} 

\section{Theory}

\subsection{Microscopic continuum dynamics}

The time evolution of a system of $N$ identical, point-like classical particles interacting via the isotropic pair potential $u(r)\equiv\varepsilon \, \tilde{u}(r/\sigma)$ (where $\epsilon$ and $\sigma$ are the fundamental energy and length scales, respectively) is governed by the canonical equations $\dot{\mathbf{r}}_i = \partial_{\mathbf{p}_i}\mathcal{H}$ and $\dot{\mathbf{p}}_i = - \partial_{\mathbf{r}_i} \mathcal{H}$, where $\mathcal{H}=\sum_{i} \frac{|\mathbf{p}_i|^2}{2\,m} + \frac{1}{2}\sum_{i,j}u(|\mathbf{r}_i-\mathbf{r}_j|)$ is the Hamiltonian of the system, $\mathbf{r}_i(t)$ and $\mathbf{p}_i(t)$ are the position and momentum of particle $i$, respectively, and $m$ is the particle mass. To re-cast the canonical equations in continuum form, we introduce the \textit{microscopic} mass and momentum densities
\begin{eqnarray}
\label{eq:Mrho}\hat{\rho}(\mathbf{r},t) &\equiv& \sum_i m \, \delta[\mathbf{r}-\mathbf{r}_i(t)] \enskip ;\\
\label{eq:Mmom}\hat{\mathbf{g}}(\mathbf{r},t) &\equiv& \sum_i \mathbf{p}_i(t) \, \delta[\mathbf{r}-\mathbf{r}_i(t)] \enskip ,
\end{eqnarray}
respectively, where $\delta(\mathbf{r})=\delta(x)\delta(y)\delta(z)$ is the three-dimensional Dirac-delta distribution. Assuming that the Fourier Transform of the pair potential exists, following the methodology of Zaccarelli et al. \cite{Zaccarelli_2002,Archer_2006,T_th_2020}, then non-dimensionalising the equations by introducing the dimensional unit length $\lambda\equiv\bar{l}\,\sigma$, time $\tau\equiv\lambda \sqrt{m/\varepsilon}$, and mass density $\rho_0$ (where $\bar{l}=\bar{n}^{-1/3}$ with $\bar{n}=\rho_0(\sigma^3/m)$, and $\rho_0$ is the dimensional average mass density of the system) result in the following dimensionless microscopic continuum equations (See Appendix A for the detailed derivation):
\begin{eqnarray}
\label{Mcont} \partial_t \hat{\rho} + \nabla \cdot \hat{\mathbf{g}} &=& 0 \enskip ; \enskip \\
\label{Mmomd} \partial_t \hat{\mathbf{g}}+\nabla\cdot\hat{\mathbb{\mathbb{K}}} &=&-\hat{\rho}\,\nabla( v * \hat{\rho} ) \enskip ;\\
\label{Mclosure}
\hat{\rho}\,\hat{\mathbb{K}} &=& \hat{\mathbf{g}} \otimes \hat{\mathbf{g}} \enskip ,
\end{eqnarray}
where $v(r)=\tilde{u}(\bar{l}\,r)$, while the symbols $*$ and $\otimes$ stand for convolution in space and dyadic product, respectively. Furthermore, $\hat{\rho}(\mathbf{r},t)=\sum_i \delta[\mathbf{r}-\mathbf{r}_i(t)]$, $\hat{\mathbf{g}}(\mathbf{r},t)=\sum_i \mathbf{v}_i(t) \delta[\mathbf{r}-\mathbf{r}_i(t)]$, and $\hat{\mathbb{K}}(\mathbf{r},t)=\sum_i \mathbf{v}_i(t)\otimes\mathbf{v}_i(t) \delta[\mathbf{r}-\mathbf{r}_i(t)]$ are the dimensionless mass, momentum and kinetic stress densities, respectively, where $\mathbf{r}_i(t)$ is the dimensionless particle position and $\mathbf{v}_i(t)$ the dimensionless particle velocity. {It is important to note that since the Hamiltonian particle dynamics is closed for the positions and momenta, the exact microscopic continuum equations are also closed for the mass and momentum densities, and there is no need to introduce further densities at this stage.} \rp{However, the practical value of Eqs. (3)-(5) is strongly limited for the following reason. Since the microscopic densities constructed by using the solution of the Hamiltonian equations satisfy Eqs. (\ref{Mcont})-(\ref{Mclosure}) only in the Fourier sense, finding this solution of Eqs. (\ref{Mcont})-(\ref{Mclosure}) is equivalent to solving the Hamiltonian particle dynamics.} 

Finally, it is worth to mention that the right-hand side of Eq. (\ref{Mmomd}) is the microscopic \rp{prototype} of the local Gibbs-Duhem relation known from Dynamical Density Functional Theory \cite{doi:10.1063/1.4913636} or phenomenological continuum models \cite{Wheeler08081997,doi:10.1146/annurev.fluid.30.1.139,Anderson2000175}. \rp{This can be shown by expressing the dimensionless potential energy of the system in terms of the microscopic mass density (see Appendix A for details):}
\begin{equation}
V = \frac{1}{2}\sum_{i,j}v(|\mathbf{r}_i-\mathbf{r}_j|) = \frac{1}{2} \int d\mathbf{r} \left\{ \hat{\rho}\,\left( v * \hat{\rho} \right) \right\} \enskip .
\end{equation} 
The first functional derivative of the potential energy with respect to the mass density then reads $\delta V/\delta \hat{\rho}=v * \hat{\rho}$, and therefore Eq. (\ref{Mmomd}) can be re-written as 
\begin{equation}
\label{response5}
\partial_t \hat{\mathbf{g}} + \nabla \cdot \hat{\mathbb{K}} = -\hat{\rho} \, \nabla \frac{\delta V}{\delta \hat{\rho}} \enskip .
\end{equation} 
It has recently been shown that the Navier-Stokes equation can be derived in the framework of statistical mechanics by coarse-graining Eq. (\ref{response5}) \cite{T_th_2020}.

\subsection{\rp{An exact scaling limit}}

To address the macroscopic-scale behaviour of the system, one can utilise an equivalence transformation \rp{between the solutions of Eqs. (\ref{Mcont})-(\ref{Mclosure}) for two different systems as follows.} The solution of the equations for particle mass $m_\kappa \equiv \kappa^3 m$, pair potential $u_\kappa(r) \equiv \kappa^{3}u(r/\kappa)$, and dimensional average number density $\bar{n}_0^\kappa \equiv \kappa^{-3} \bar{n}_0$ (called the $\kappa$-system henceforth) can be expressed as $\chi_\kappa(\mathbf{r},t) = \hat{\chi}(\mathbf{r}/\kappa,t/\kappa)$ (see Appendix B), where $\hat{\chi}(\mathbf{r},t)$ stands for the solution in the original system with $\chi=\rho,\mathbf{g},\mathbb{K}$. {The scaling relation
\begin{equation}
\hat{\chi}(\mathbf{r},t) = \chi_\kappa(\kappa\,\mathbf{r},\kappa\,t) 
\end{equation}
indicates that the $O(1)$-scale dynamics in the $\kappa$-system corresponds to the $O(1/\kappa)$-scale dynamics in the original one. Consequently,} the macroscopic dynamics of the original system can then be studied by considering the $O(1)$-scale dynamics in the $\kappa \to 0$ system. Assuming that the \textit{macroscopic} densities
\begin{equation}
\rho(\mathbf{r},t)\equiv \lim_{\kappa \to 0}\rho_\kappa(\mathbf{r},t) \quad \text{and} \quad \mathbf{g}(\mathbf{r},t)\equiv \lim_{\kappa \to 0}\mathbf{g}_\kappa(\mathbf{r},t)
\end{equation} 
are \textit{bounded functions} (instead of sums of Dirac-delta distributions), the kinetic stress can be expressed from the closure relation for $\rho(\mathbf{r},t)>0$ as: $\mathbb{K} = (\mathbf{g}\otimes\mathbf{g})/\rho$ for $\rho(\mathbf{r}) > 0$, where $\mathbb{K}(\mathbf{r},t)=\lim_{\kappa\to 0}\hat{\mathbb{K}}_\kappa(\mathbf{r},t)$, and the dynamical equations for the $\kappa \to 0$ system then read (see Appendix B): $\partial_t \rho + \nabla \cdot (\rho\,\mathbf{v}) = 0$ and $\partial_t \mathbf{v} + \mathbf{v} \cdot \nabla \mathbf{v} = - \nabla(v_0 * \rho)$, where $\mathbf{v}(\mathbf{r},t)=\mathbf{g}(\mathbf{r},t)/\rho(\mathbf{r},t)$ is the velocity field, while $v_0(r) = \bar{n}\,\lim_{\kappa \to 0}\left[ \kappa^{-3}\tilde{u}(r/\kappa) \right]$ is the scaling limit of the pair potential. For some common pair potentials listed in Table 1, it can be shown that
\begin{equation}
\label{potlimit}
v_0(r) =\bar{n}\,a_0\,\delta(\mathbf{r}) \enskip .
\end{equation}
\begin{table}
\centering
\caption{Macroscopic limit for the collision, Yukawa, screened inverse power, and the Hartree-Fock dispersion B potentials (from top to bottom), as defined by Eq. (\ref{potlimit}).}
\begin{tabular}{|c|c|}
\hline
$\tilde{u}(r)$ & $a_0$ \\
\hline
$\delta(r)$ & 1 \\
\hline
$\exp(-\alpha\,r)/r$ & $4\pi/\alpha^2$ \\
\hline
$\{[1-\exp(-\alpha\,r)]/r\}^n$ & $A_n\,\alpha ^{n-3}$ \\
\hline
$\exp(-\alpha\,r-\beta\,r^2)$ & $[\pi ^{3/2}/(2\,\beta^{5/2})] \exp[\alpha ^2/(4 \beta)]$\\
& $\times(\alpha ^2+2 \beta)\textrm{erfc}[\alpha/(2 \sqrt{\beta})]-\pi  \alpha /\beta ^2$ \\ 
\hline
\end{tabular}
\end{table}
Using Eq. (\ref{potlimit}) in the momentum equation and eliminating the group velocity $c_0^2=\bar{n}a_0$ by re-scaling time yield the following \textit{universal} exact macroscopic continuum equations to the Hamiltonian system of pair interacting classical particles:
\begin{eqnarray}
\label{MacroNS1} \partial_t \rho + \nabla\cdot(\rho\,\mathbf{v}) &=& 0 \enskip ; \\
\label{MacroNS2} \partial_t \mathbf{v} + \mathbf{v} \cdot \nabla\mathbf{v} + \nabla \rho &=& 0
\end{eqnarray}
with $\displaystyle{\int d\mathbf{r}\{\rho(\mathbf{r},t)-1\}=0}$ and $\displaystyle{\int d\mathbf{r}\{\mathbf{g}(\mathbf{r},t)\}=0}$.

\subsection{\textcolor{red}{Properties of the macroscopic densities}}

\rp{Before proceeding to the analysis of the numerical solutions of Eqs. (\ref{MacroNS1}) and (\ref{MacroNS2}), we need to know whether the scaling limits of the microscopic densities are at least bounded functions, since in this case they can be found by solving the dynamical equations directly. This is a major difference between Eqs. (\ref{Mcont})-(\ref{Mclosure}) and (\ref{MacroNS1})-(\ref{MacroNS2}). The boundedness of the macroscopic density fields can be argued as follows.} \rp{Using the standard assumption of separable scales, the dimensionless particle positions and momenta can be written as a sum of a fast and a slowly varying component} (see Figure 1):
\begin{eqnarray}
\label{DCpos} \mathbf{r}_{j(i)}(t) &=& \mathbf{r}_i^0 + \mathbf{u}(\kappa\,\mathbf{r}_i^0,\kappa\,t)/\kappa + \Delta\mathbf{r}_{j(i)}(t) \enskip ; \enskip\\
\label{DCmom} \mathbf{v}_{j(i)}(t) &=& \mathbf{w}(\kappa\,\mathbf{r}_i^0,\kappa\,t) + \Delta\mathbf{v}_{j(i)}(t) \enskip , \enskip
\end{eqnarray}
where $\mathbf{r}_i^0 \in \mathbb{Z}^3$ spans a uniform grid with unit grid spacing, $\mathbf{u}(\mathbf{r},t)$ and $\mathbf{w}(\mathbf{r},t)$ are sufficiently smooth zero-average vector fields describing spatio-temporal variations of the mass and momentum densities on $O(1/\kappa)$ scale, respectively, while $\Delta\mathbf{r}_{j(i)}(t)$ and $\Delta\mathbf{v}_{j(i)}(t)$ carry the microscopic (or thermal-scale) details.
\begin{figure}
\centering
\includegraphics[width=0.666\linewidth]{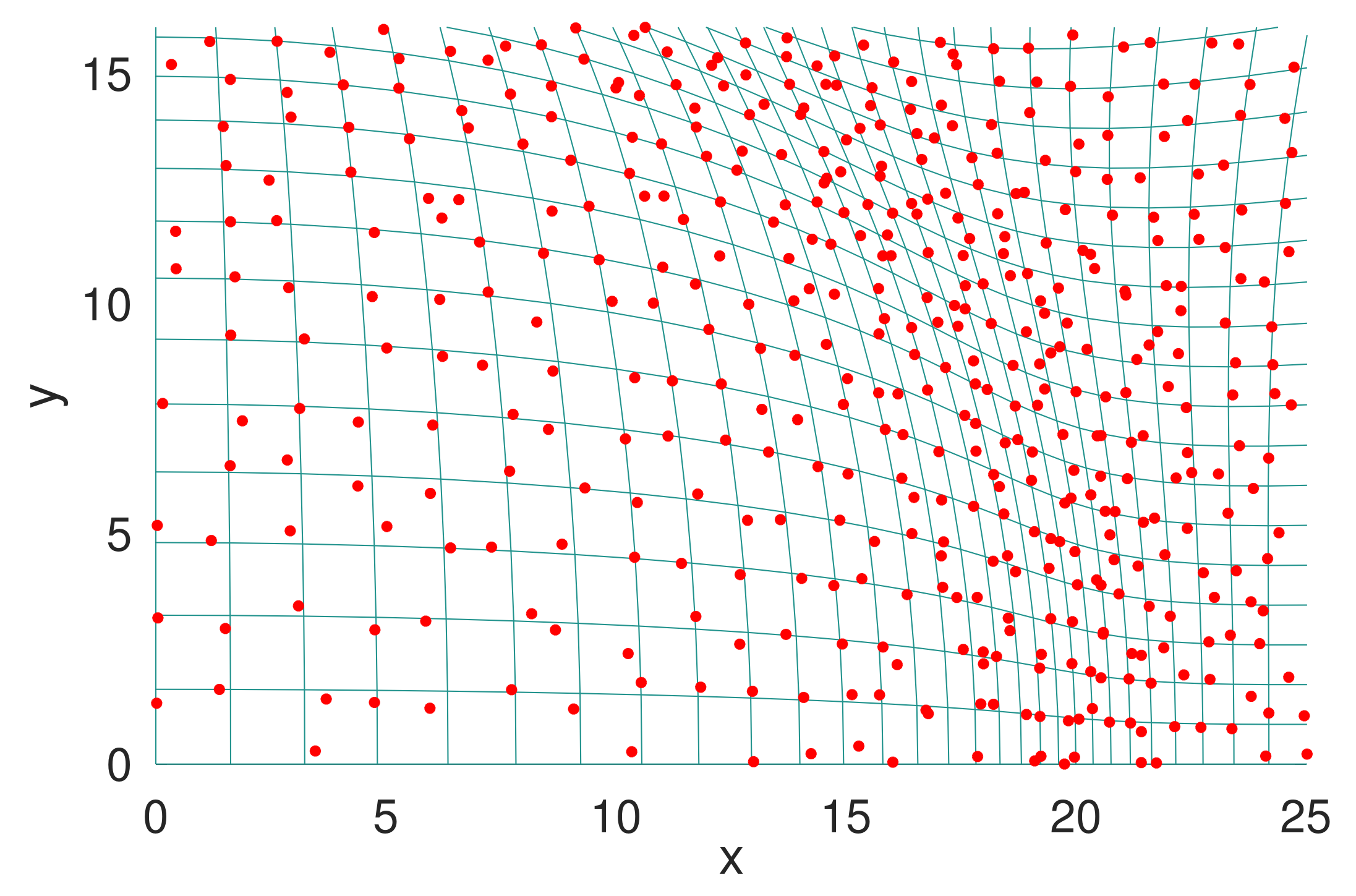}
\caption{Schematic illustration of Eqs. (13) and (14) in a two-dimensional
system. The component of the density varying on length scale $\kappa$ is illustrated by the smoothly deformed grid, while the actual particle positions are indicated by the off-grid dots.}
\end{figure}
Furthermore, $j(i)$ is an instantaneous map between the lattice sites (indexed by $i \in \mathbb{N}$) and the particles (indexed by $j \in \mathbb{N}$). Considering only one spatial dimension, {substituting Eqs. (\ref{DCpos}) and (\ref{DCmom}) into Eq. (\ref{eq:Mrho}) yields (see Appendix C for a detailed derivation):
\begin{equation}
\label{eq:macrolimit1} \rho(x,t) = 1+\sum_{p=1}^\infty (1/p!)(-\partial_x)^p [u^p(x,t)] \enskip .
\end{equation}
{A similar calculation can be performed for the momentum density, thus yielding:}
\begin{equation}
\label{eq:macrolimit2} g(x,t) = \sum_{p=0}^\infty (1/p!)(-\partial_x)^p [\omega(x,t)u^p(x,t)] \enskip ,
\end{equation}
where $\omega(x,t)=w(x,t)+\xi(x,t)$ with 
\begin{equation}
\label{thermo}
\xi(x,t)=\lim_{\kappa\to 0}\Delta v_{j(\lfloor x/\kappa \rfloor)}(t/\kappa)
\end{equation} 
{with $\lfloor x \rfloor$ denoting the floor of $x$.} Eqs. (\ref{eq:macrolimit1}) and (\ref{eq:macrolimit2}) indicate that $\rho(x,t)$ and $g(x,t)$ can be bounded functions. (The boundedness of $g(x,t)$ in case of $v_i(t) \neq 0$ can be argued on the basis of the invariance of the dimensionless particle velocities in the equivalent systems.) \rp{Finally we mention that differentiability of $\rho(\mathbf{r},t)$ and $\mathbf{g}(\mathbf{r},t)$ is not required: In case the macroscopic densities are only weak solutions to Eqs. (\ref{MacroNS1})-(\ref{MacroNS2}), they can still be found numerically.}

Beyond assuming the boundedness of the macroscopic densities, an important feature of Eq. (\ref{eq:macrolimit2}) is that $\Delta \mathbf{v}_i(t)$ \rp{is present} in $\omega(x,t)$, thus suggesting that the macroscopic momentum field \rp{incorporates} the temperature \rp{(in form of a rapidly varying function, see Eq. (\ref{thermo})).} This can be \rp{further argued} by calculating the instantaneous dimensionless temperature of the $\kappa=1$ system: $\tilde{T}(t) \equiv T(t)/T_0 = \frac{1}{3} \langle \textrm{Tr}\,\hat{\mathbb{K}}(\mathbf{r},t) \rangle$, where $\langle.\rangle$ stands for spatial average, $T(t) \equiv (3 k_B N\,m)^{-1} \sum_i |\mathbf{p}_i|^2$ is the dimensional temperature, and $T_0=\epsilon/k_B$ the temperature scale. Since the spatial averages of the corresponding microscopic densities coincide in the equivalent systems, the temperature can be expressed in terms of the macroscopic fields in the $\kappa \to 0$ system as:
\begin{equation}
\label{temperature}
\Theta(t) \equiv 3\,\tilde{T}(t)/(a_0\,\bar{n}) = \left< \rho(\mathbf{r},t)\,|\mathbf{v}(\mathbf{r},t)|^2 \right> \enskip .
\end{equation}
\rp{Eqs. (\ref{eq:macrolimit2}), (\ref{thermo}) and (\ref{temperature}) indicate that only those solutions of Eqs. (\ref{MacroNS1})-(\ref{MacroNS2}) represent scaling limits of microscopic densities at finite temperatures, in which the velocity field contains a rapidly varying thermal component. This means that} Eqs. (\ref{MacroNS1}) and (\ref{MacroNS2}) are cardinally different from that of the  \rp{coarse-grained} hydrodynamic equations, which only describe the spatio-temporal evolution of slowly varying fields. In \rp{the present work, however,} the derivation of Eqs. (\ref{MacroNS1}) and (\ref{MacroNS2}) from the many-body dynamics is exact, and therefore they are expected to preserve the microscopic details \rp{in the form of rapidly varying components}. At this point, it is important to note that this idea is not completely \rp{novel}. Recent results indicate the \rp{emergence of rapidly varying solutions (called ``tygers'')} of the Burgers equation \cite{Ray2011,doi:10.1098/rspa.2016.0585,RAY2015}, which \rp{- in the context of the present work -} simply describe the macroscopic\rp{-scale} dynamics of non-interacting systems.

\rp{Summarising the above argumentation, the main advantage of Eqs. (\ref{MacroNS1}) and (\ref{MacroNS2}) is that the emergence of macroscopic-scale pseudo time-irreversibility due to microscopic initial conditions can be studied by solving the macroscopic dynamical equations for suitable initial conditions. To find suitable initial conditions for the momentum density, first we consider a Hamiltonian system with initial conditions $\mathbf{r}_{j(i)}(0) = \mathbf{r}_i^0$ (lattice sites on a uniform grid with grid spacing 1) and $\mathbf{v}_{j(i)}(0) = \vec{\mathcal{G}}_T$ (random vectors according to the Maxwell-Boltzmann distribution). The scaling limit of the initial microscopic densities are exact and read: $\rho(\mathbf{r},0)=1$ and $\mathbf{g}(\mathbf{r},0)=\vec{\mathcal{G}}_T(\mathbf{r})$ (Gaussian random vector field). Since both the macroscopic dynamical equations and the scaling limit of the initial conditions are exact, the solution of the macroscopic equations are expected to be the scaling limits of the microscopic mass and momentum densities in a Hamiltonian system starting from the corresponding microscopic initial conditions. In addition, the Markovian component (Hypothesis of Molecular Chaos restricted to the initial conditions) is incorporated in the dynamics only on the level of the initial conditions, the effect of random initial conditions on the macroscopic-scale spatio-temporal evolution of the system can be studied. Finally, if the numerical solution of the macroscopic dynamical equations remains bounded, the assumption of boundedness of $\rho(\mathbf{r},t)$ and $\mathbf{g}(\mathbf{r},t)$ is further supported.}

\section{Numerical results}

\subsection{Equilibration and equipartition}

The first step in studying the emergence of pseudo time-irreversibility due to random initial conditions is the characterisation of thermodynamic equilibrium.
Using the Fourier representation of the velocity field for a finite-size and discretised system, the temperature reads: $\Theta(t) \approx \sum_{\mathbf{k}}|\mathbf{v}_\mathbf{k}(t)|^2$ for $\rho(\mathbf{r},t)\approx 1$, which indicates the hypothesis of spectral equipartition:}
\begin{equation}
\label{eq:sequip}
\Theta_\infty \equiv \lim_{t \to \infty} \frac{1}{t}\int_0^t d\tau\,\Theta(\tau) = N_k V^2_\infty \enskip , 
\end{equation}
where $V^2_\infty = \frac{1}{t} \int_0^t d\tau |\tilde{\mathbf{v}}_\mathbf{k}(\tau)|^2$ (constant), and $N_k$ is the number of Fourier modes. \rp{Eq. (\ref{eq:sequip}) expresses that the momentum is expected to spread evenly across the Fourier modes in equilibrium.} To validate the hypothesis, numerical simulations were performed for $\Theta_0 = 5 \times 10^{-4}$ initial temperature, mimicking liquid Argon below its boiling point \cite{doi:10.1063/1.458165,doi:10.1063/1.466051}. To exploit the conservation of mass and momentum, the numerical implementation of Eqs. (\ref{MacroNS1}) and (\ref{MacroNS2}) was done in the mass-momentum formalism $\partial_t \rho + \nabla\cdot\mathbf{g}=0$ and $\partial_t \mathbf{g} + \nabla\cdot[(\mathbf{g}\otimes\mathbf{g})/\rho+\rho\nabla\rho]=0$ by using a flux-consistent finite-volume scheme and forward Euler time discretisation \cite{chandrashekar_2013} on a two-dimensional uniform grid with grid size $N=1024$, grid spacing $h=1$ and time step $\Delta=10^{-4}$. We applied periodic boundary conditions, thus resulting in a discrete Fourier space with resolution $\Delta k=2\,\pi/1024$. \rp{We note that Eqs. (\ref{MacroNS1}) and (\ref{MacroNS2}) are scale-free in both space and time, and therefore no qualitatively different behaviour is expected to be observed for different choices of $h$, $\Delta t$ and $N$. Also, the lack of characteristic scales indicates the self-similarity of the solution, and therefore uncorrelated spatio-temporal (pseudo) random fields can represent stable weak solutions of the dynamical equations. Since the Finite Volume numerical scheme is a simple Finite Element scheme, these weak solutions should emerge in the numerical simulations.}

\begin{figure}
\centering
\includegraphics[width=0.333\linewidth]{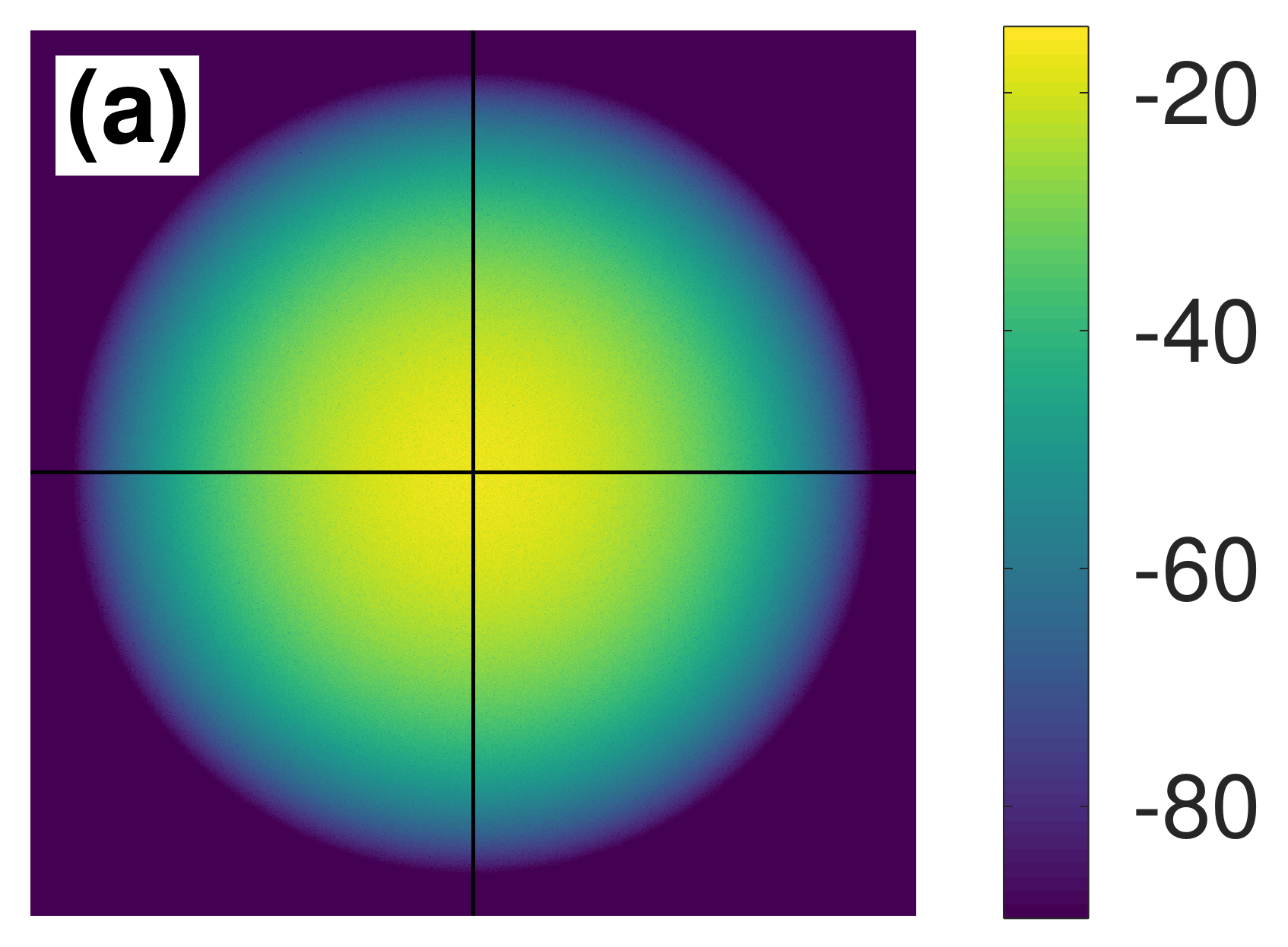}
\includegraphics[width=0.333\linewidth]{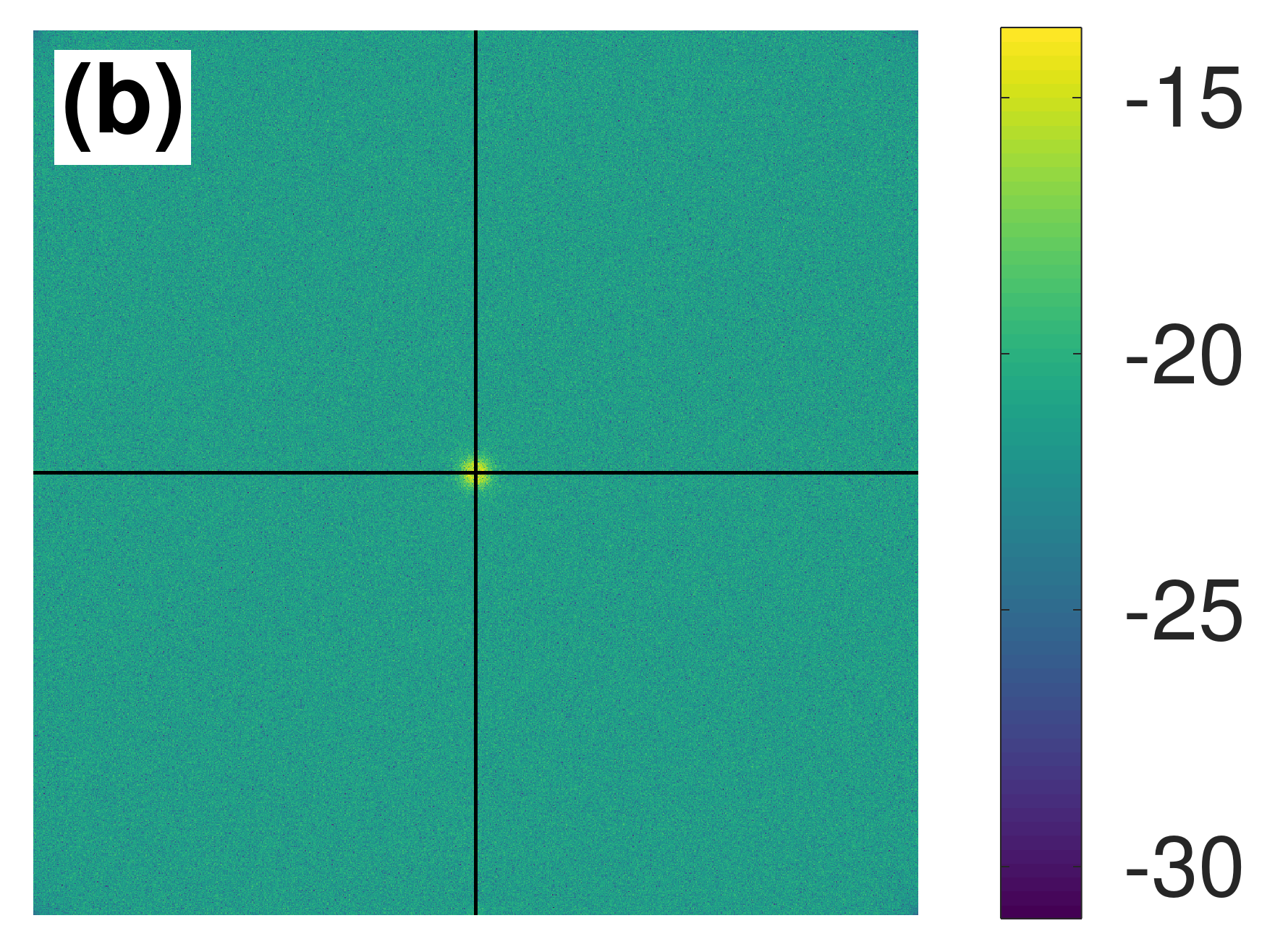}\\
\includegraphics[width=0.666\linewidth]{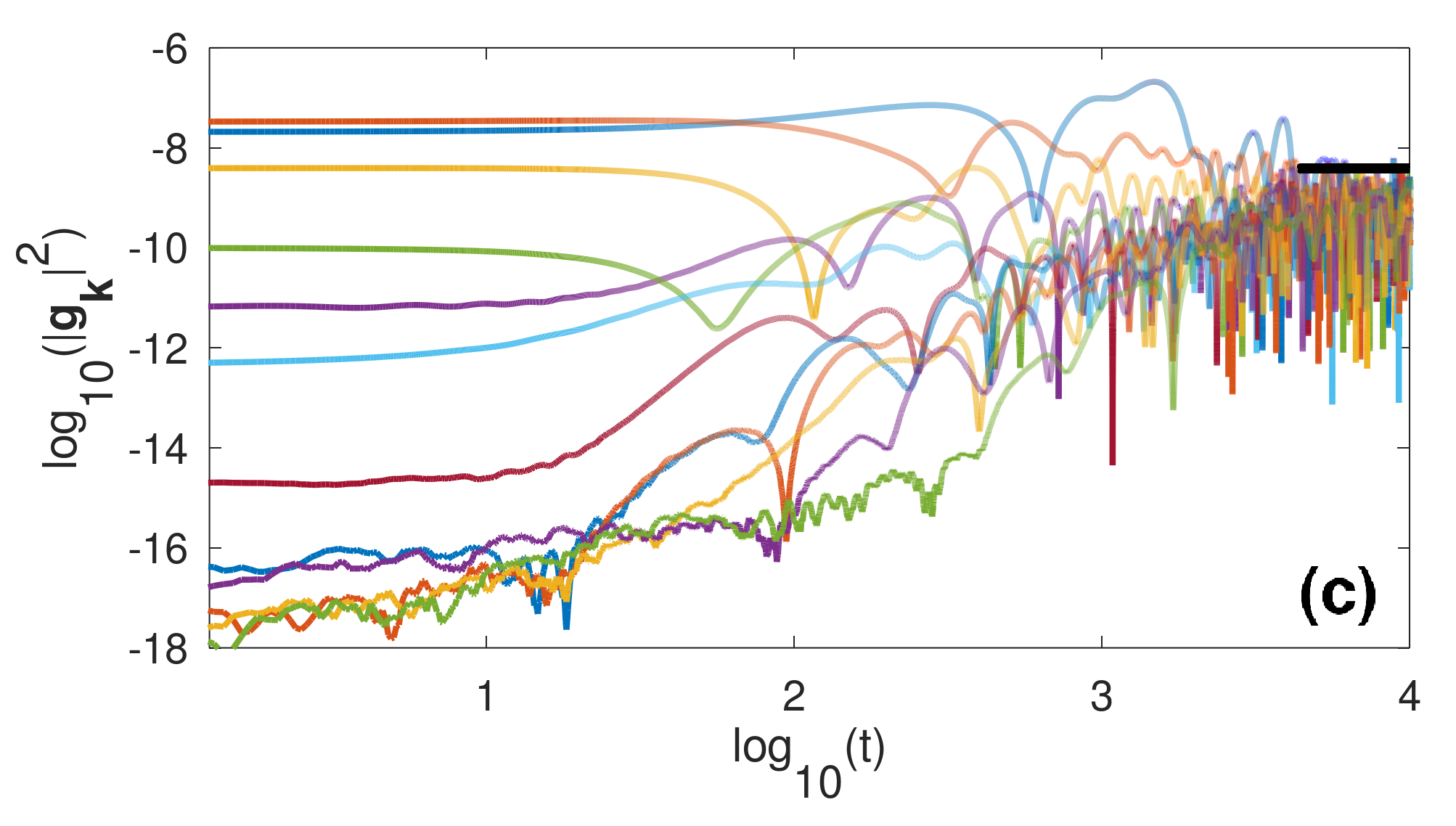}
\caption{Fourier amplitudes (logarithmic scale) of the momentum density in the  range $[-\pi,\pi]\times[-\pi,\pi]$ at (a) $t=0$ and (b) $t=10^{4}$; (c) Time dependence of spectral momentum components at $\mathbf{k}=(40\,\Delta k\,n,0)$, where $n=1,2,\dots,12$ (from top to bottom, respectively).}
\end{figure}
In accordance with Section 2.3, the initial conditions were chosen as $\rho(\mathbf{r},0)=1$ and $\tilde{\mathbf{g}}_{\mathbf{k}}(0) = A\,f(k)\,(\mathbb{I} - \mathbf{n}_{\mathbf{k}} \otimes \mathbf{n}_{\mathbf{k}}) \cdot\vec{\xi}_{\mathbf{k}}$, where $\vec{\xi}_{\mathbf{k}}$ is an uncorrelated Gaussian random vector field in the Fourier space, $A = (\pi/N)\sqrt{6\,\Theta_0/I}$ and $I=\int_0^\pi (2\,\pi\,k\,dk) f^2(k)$. The out-of-equilibrium initial condition was assured by choosing $f(k)=\text{sinc}^{16}(k)\,\theta(\pi-k)$ (describing a macroscopically ordered momentum field by long wavelength spatial correlations). The time evolution of the system was studied for $n=10^{8}$ time steps. While the spectral components of the momentum (denoted by $|\mathbf{g}_\mathbf{k}(t)|^2$) were fast decaying around $\mathbf{k}=\mathbf{0}$ at $t=0$, by the end of the simulation the momentum had spread across the available wave numbers [see Figs. 2(a) and 2(b)]. The time evolution of the square magnitude of 12 different Fourier components of the momentum density are shown in Fig. 2(c). The figure indicates that Fourier amplitudes initially spanning an 11 orders of magnitude range ``converge'' for large times in the sense that their maxima tend to be in the same order of magnitude (indicated by the thick horizontal line in the figure). In addition, since the relative density and temperature fluctuations remained in the range of $\pm0.5\%$ and $\pm0.025\%$, respectively, the results provide evidence for the presence of thermodynamic equilibrium with spectral equipartition defined by Eq. (\ref{eq:sequip}).
\begin{figure}
\centering
\includegraphics[width=0.333\linewidth]{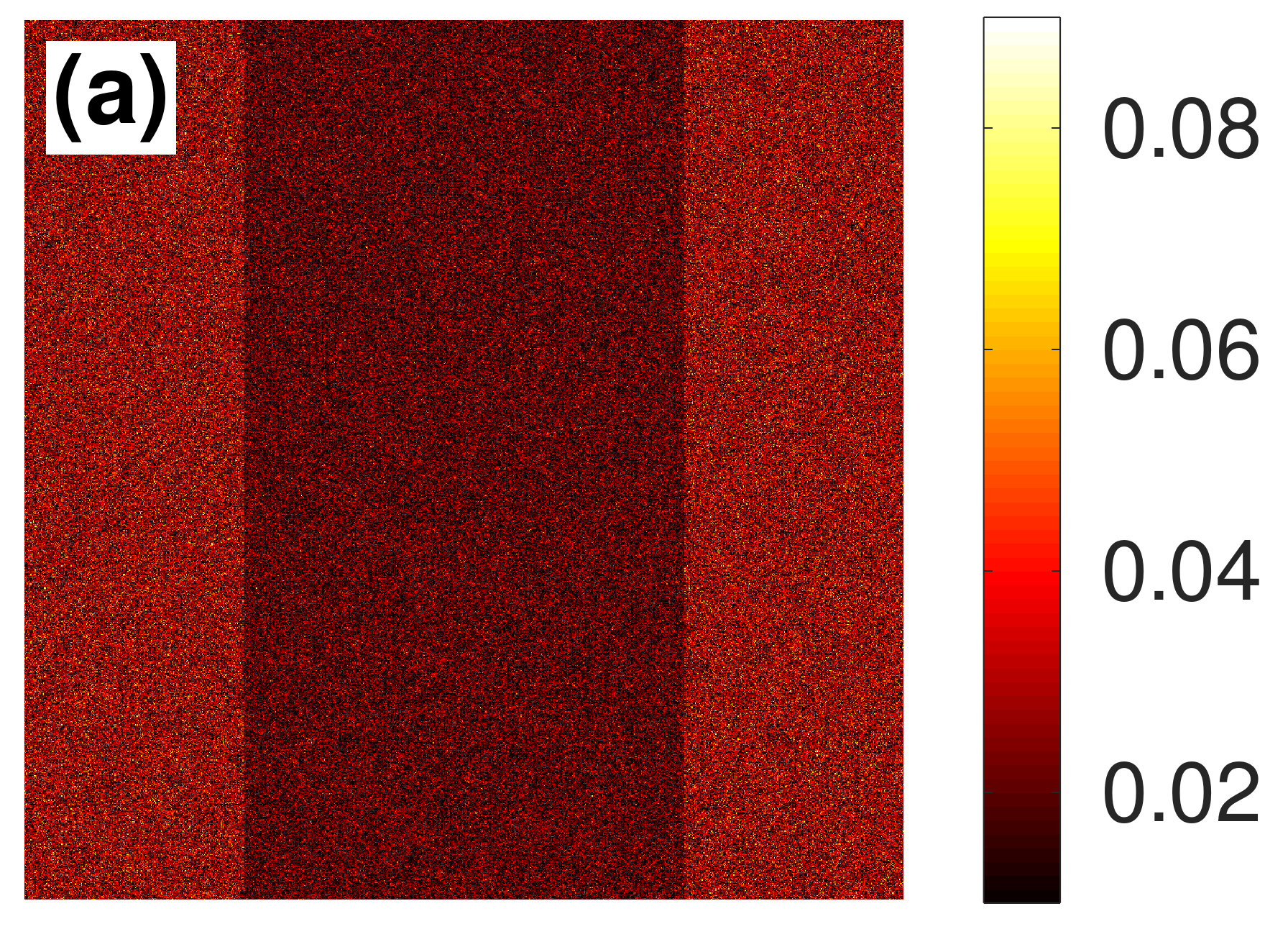}
\includegraphics[width=0.333\linewidth]{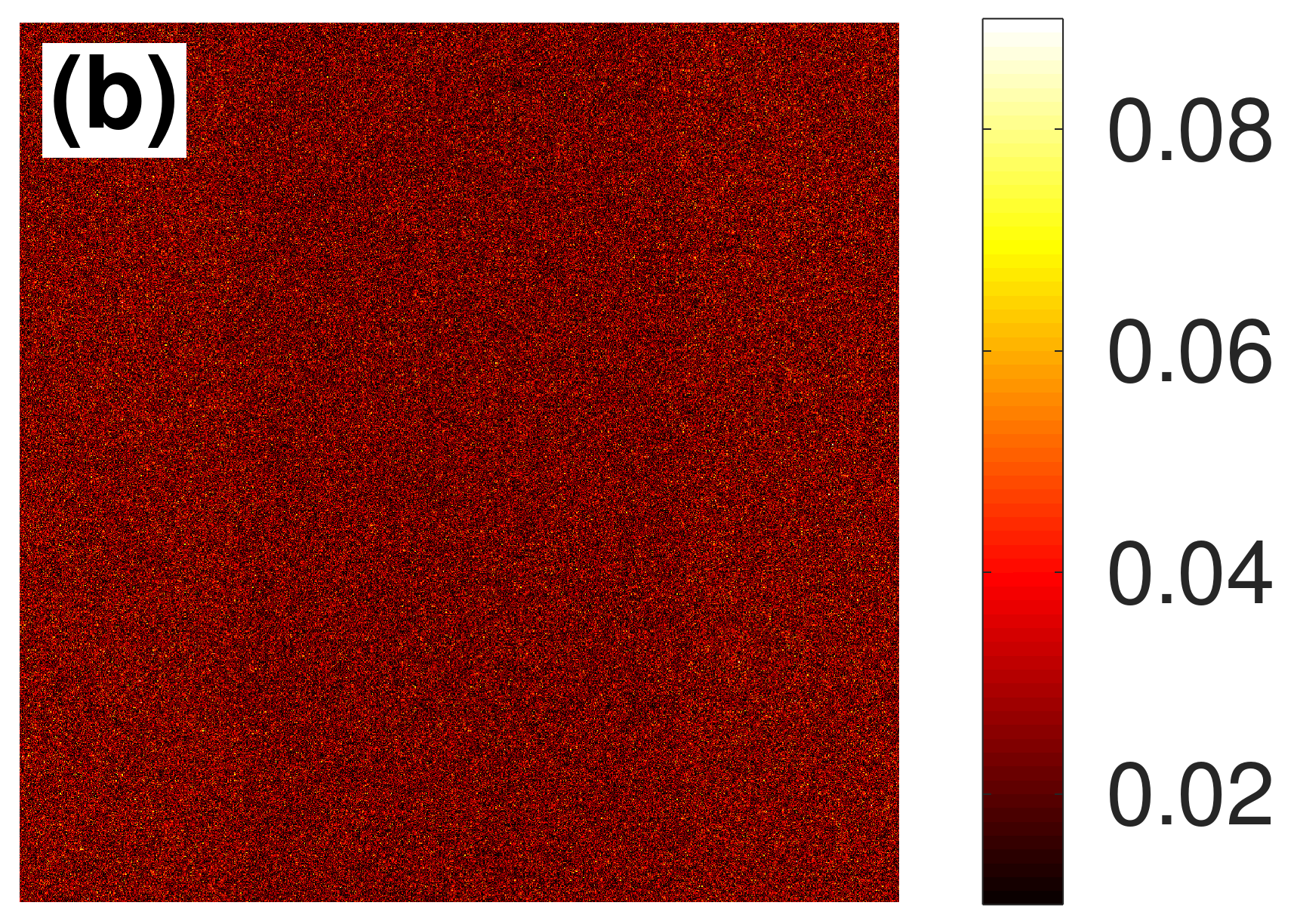}\\
\includegraphics[width=0.666\linewidth]{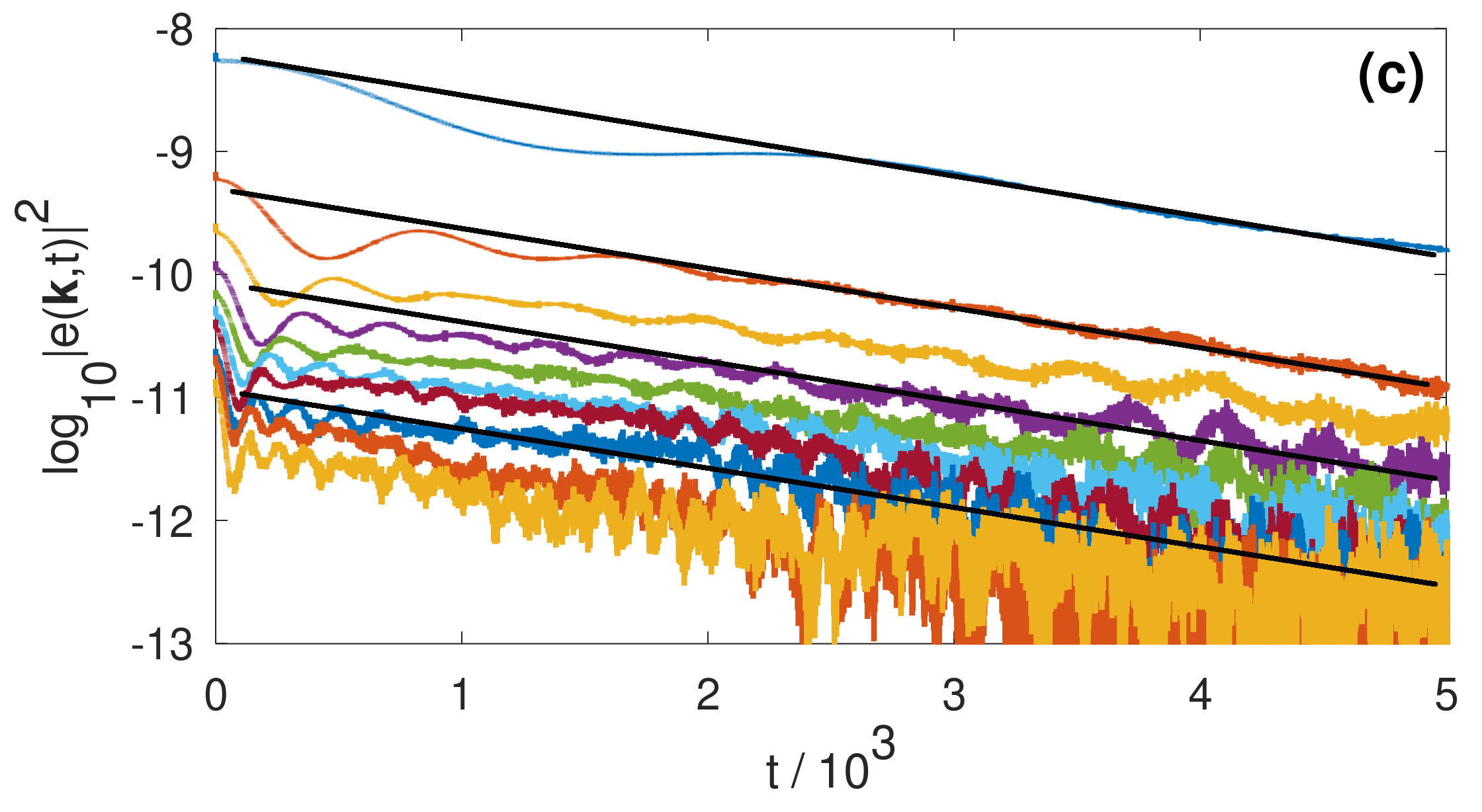}
\caption{Kinetic energy density in real space at (a) $t=0$ and (b) $t=5\times 10^3$; (c) Time evolution of the Fourier coefficients of $e(\mathbf{r},t)$ for $\mathbf{k}=(\Delta k\,n,0)$, where $n=1,3,5,\dots,19$ (from top to bottom).}
\end{figure}

\subsection{Heat conduction}

In the second numerical simulation, heat transport was investigated by bringing two equilibrated systems with temperatures $T_1=10^{-3}$ and $T_2=T_1/2$ into contact as shown in Fig 2(a). (In the equilibration process of the sub-systems a homogeneous thermostat was applied in every $10^5$ time steps to prevent the system from overheating due to Galerkin trancation.) The temperature difference between the regimes of different temperatures started to decrease gradually in an equalization process, and almost completely vanished by $t=5\times 10^3$ [see Figs. 3(b) and 3(c)], thus confirming the presence of heat condition. Nevertheless, Fig. 3(d) shows the time evolution of the magnitude of some Fourier coefficients of the kinetic energy density $e(\mathbf{r},t)=(1/2)\rho(\mathbf{r},t)|\mathbf{v}(\mathbf{r},t)|^2$ (or local temperature). The figure indicates exponential relaxation of the Fourier modes with roughly the same relaxation time, which suggests a pseudo-irreversible process with a local relaxation dynamics $\partial_t \Theta(\mathbf{r},t) = -\alpha[\Theta(\mathbf{r},t)-\Theta_\infty]$ rather than diffusion emerging from the phenomenological Fourier's law.
\begin{figure}
\centering
\includegraphics[width=0.333\linewidth]{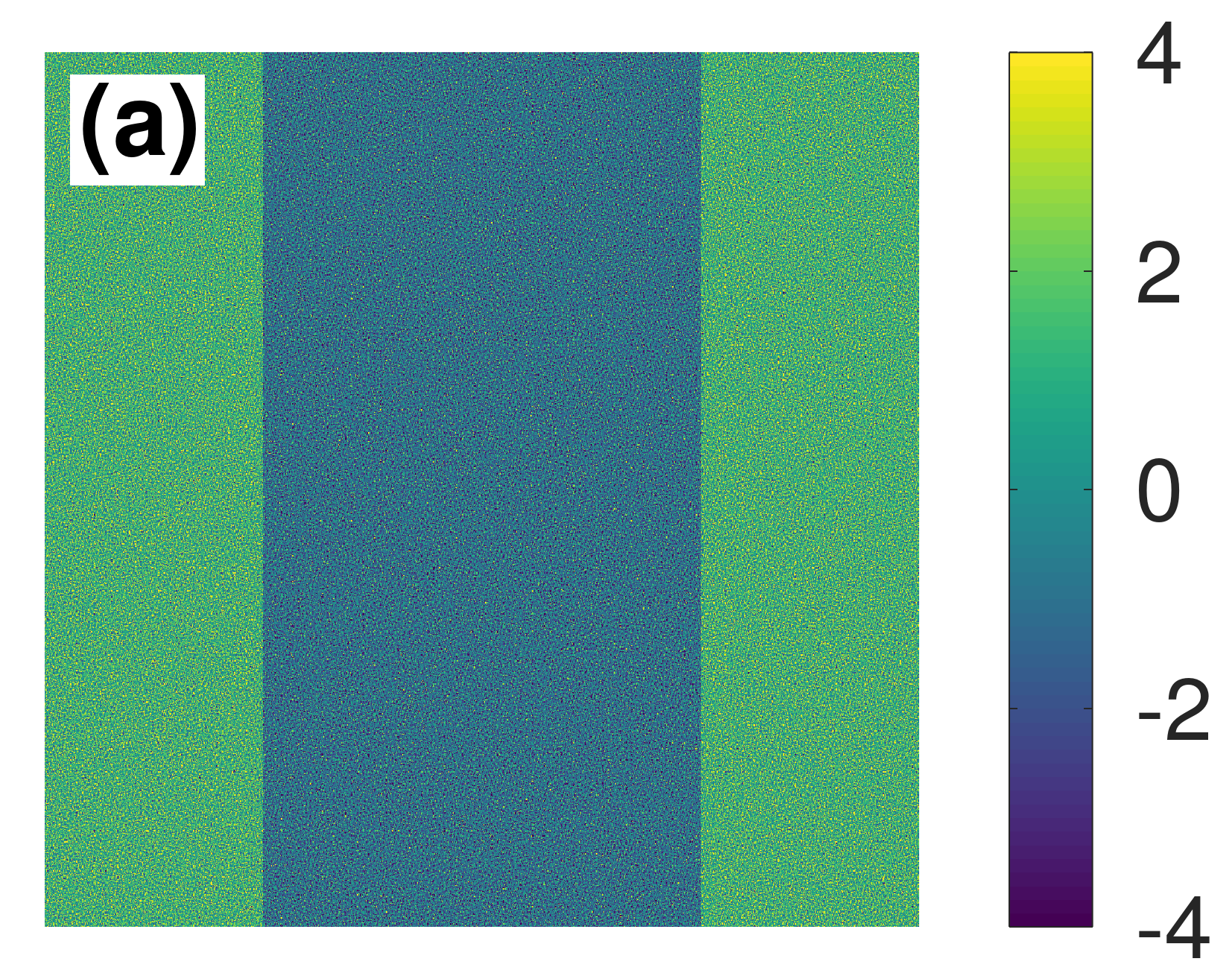}
\includegraphics[width=0.333\linewidth]{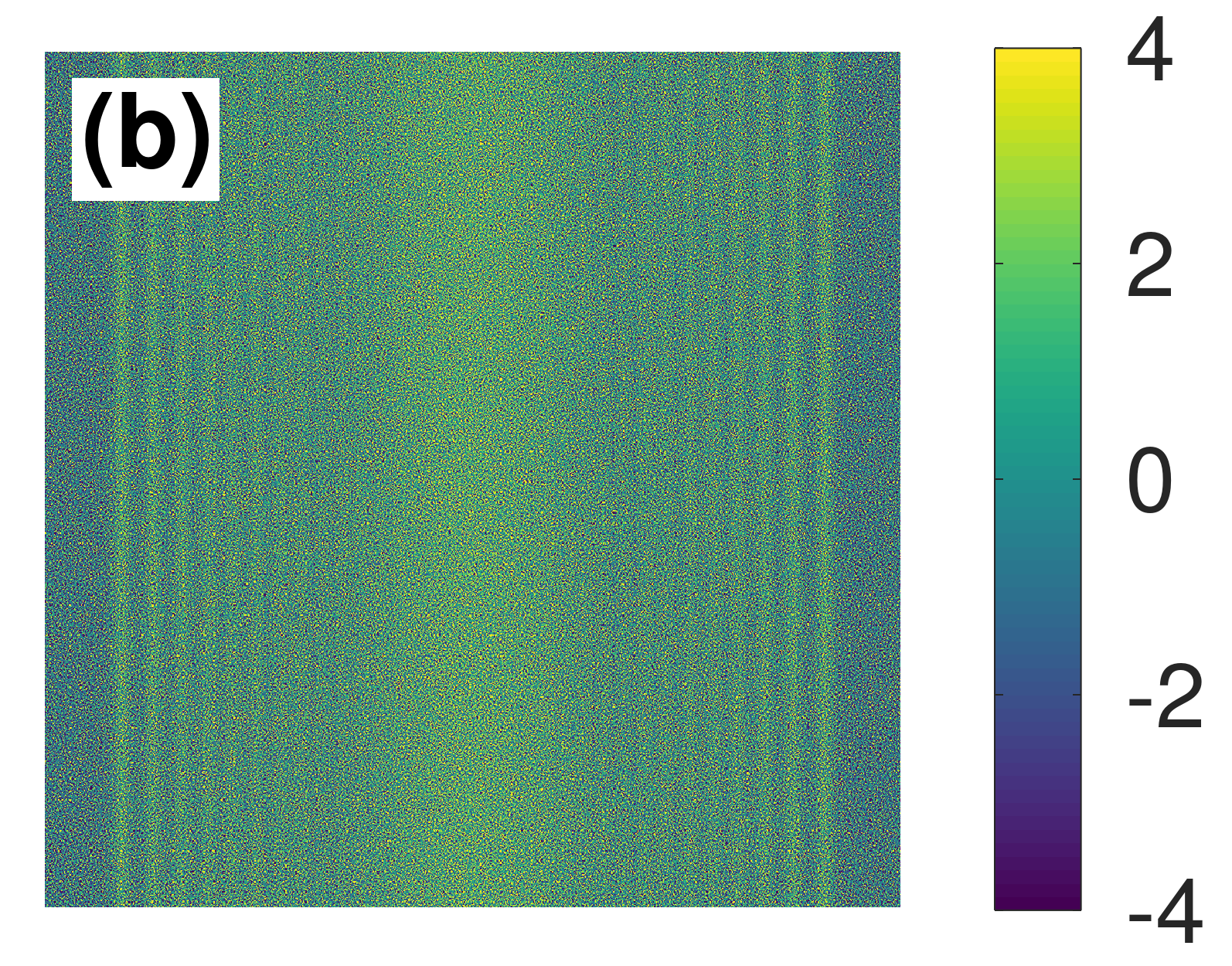}\\
\includegraphics[width=0.333\linewidth]{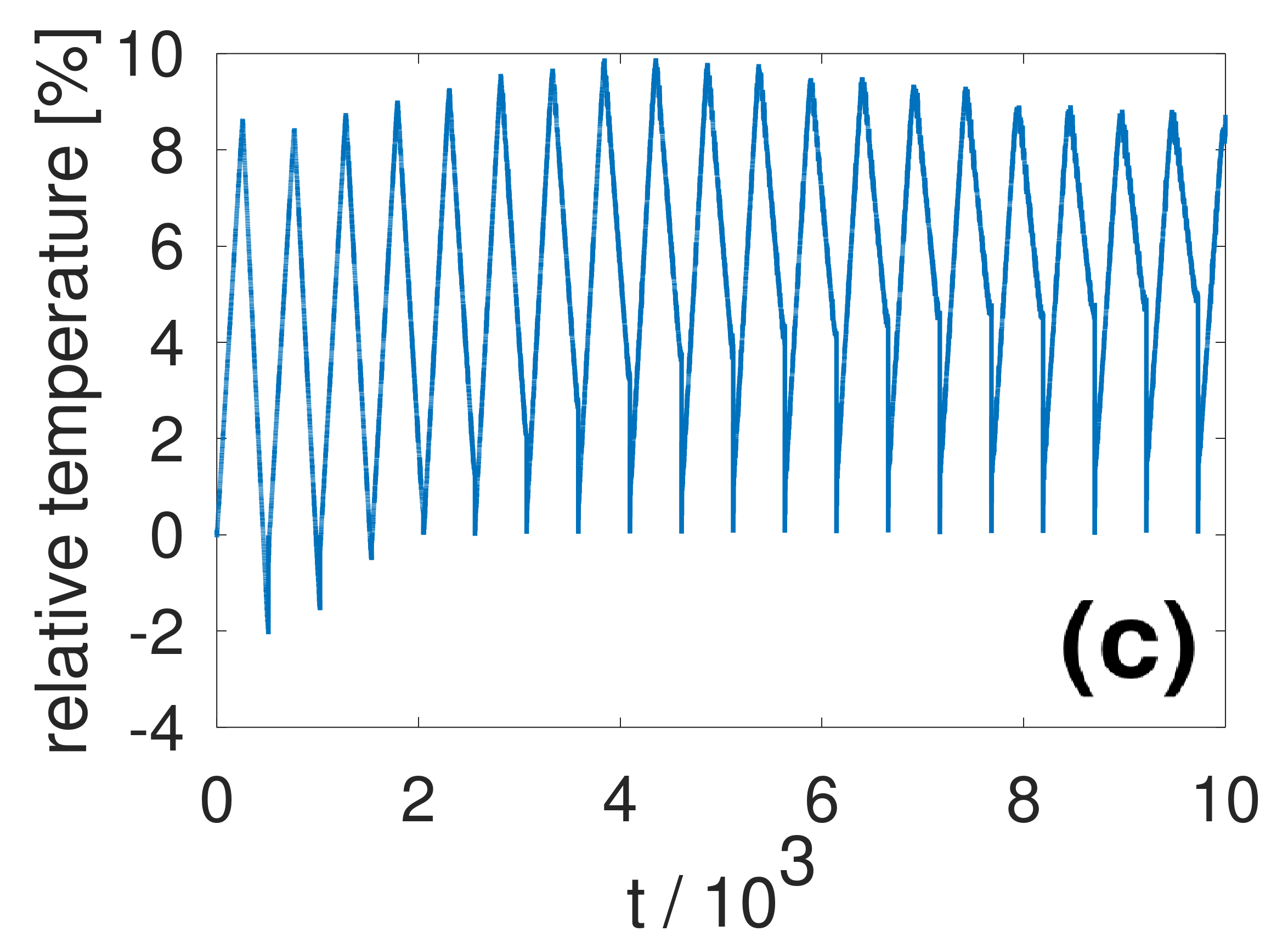}
\includegraphics[width=0.333\linewidth]{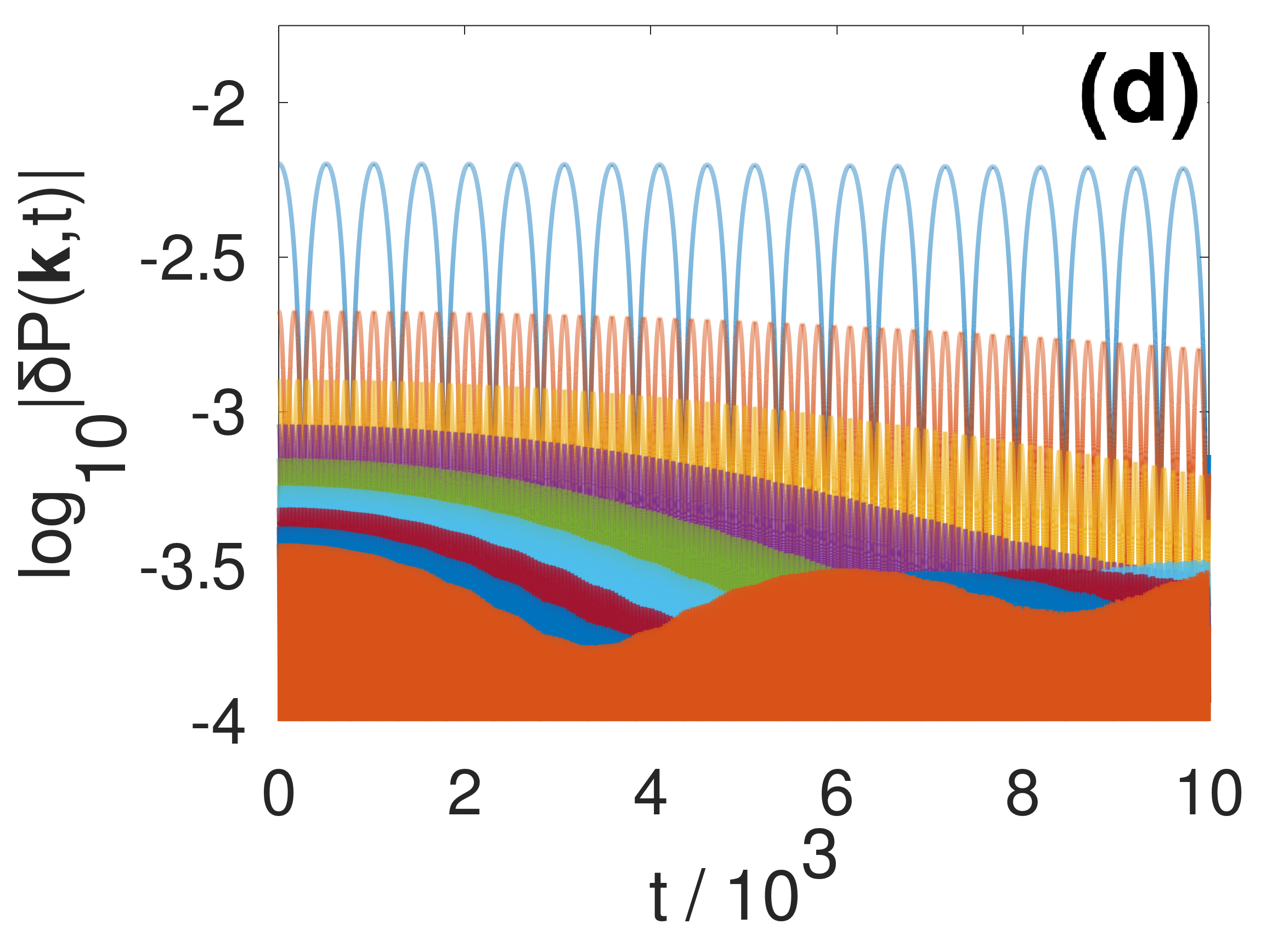}
\caption{Relative density difference at (a) $t=0$ and (b) $t=10^4$; (c) Time evolution of the relative temperature $100[\Theta(t)/\Theta_0-1]$; (d) Time evolution of the Fourier amplitudes of the density for wave numbers $\mathbf{k_0}=(\Delta k\,n,0)$ for $n=1,3,\dots,19$ (from top to bottom at $t=0$).}
\end{figure}

\subsection{Viscous flow}

In our last numerical experiment, viscous momentum transport was studied by bringing two equilibrated systems of average densities $\rho_1=0.99$ and $\rho_2=1.01$ into contact at $\Theta_0=10^{-3}/3$ [see Fig. 4(a)]. To prevent the system from overheating, we applied a homogeneous thermostat in every $N_T=10^4$ steps, in which the temperature was reduced by a $0.01\%$. In addition, the temperature was brought back to $\Theta_0$ in every $N_P=512/\Delta t$ time steps. With these techniques, the temperature was kept within the relative range $\pm 6\%$ [see Fig. 4(c)]. As indicated by Fig. 4(c), the temperature started to oscillate at angular frequency $\omega_T=2\,\pi/512$, which more resembles the FPUT-type recurrence rather than pseudo-irreversibility emerging from a diverging Poincar\'e recurrence time. The decay of the macroscopic order in the density field is indicated by Figs. 4(b) and 4(d). As shown in Fig. 4(d), the long wavelength Fourier amplitudes of the density evolved in the fashion of $|\delta P(k,t)| \propto \cos^2(\omega_k t)\{1+\alpha\,[\sin(\Omega_k t)/(\Omega_k t)]^p\}$ (where $\omega_k = \Delta k\,\sqrt{2^{n+1}}$ for $k=\Delta k\,n$, $n=1,3,5,\dots$), which indicates the decay of the initial macroscopic order in the studied time interval. However, despite the relatively small difference of the average densities of the two initial sub-systems, the dynamics of the relaxation is different compared to the diffusive relaxation provided by the linearisation of the Navier-Stokes equations, which provides $|\delta P(k,t)|^2 \propto e^{-\mu\,k^2\,t}$ for a viscous one-dimensional ideal gas (where $\mu$ is the viscosity).\\

\section{Summary}

\rp{In this paper, emerging macroscopic-scale pseudo-time irreversibility due to random microscopic initial conditions was studied in the closed Hamiltonian system of pair interacting particles. First, exact continuum equations were developed to the Hamiltonian particle system in such a scaling limit, in which - contrary to coarse-grained models - the thermal component of the microscopic momentum density is preserved in the form of rapid variations in the macroscopic field. Because of the presence of the thermal component in the scaling limit, the derived model can be positioned between full-resolution microscopic models and coarse-grained continuum models. Consequently, the derived continuum equations make possible to study the effect of special microscopic initial conditions on the macroscopic-scale spatio-temporal evolution of the system. It has been argued that the thermal component of the microscopic momentum density constructed on the basis of the solution of the Hamiltonian equations should scale into an uncorrelated (in both space and time) pseudo-random field. Consequently, a finite temperature system was initiated by an uncorrelated Gaussian random field on the level of initial conditions for the macroscopic momentum density. With this approach, the Markovian stochastic component (utilised in coarse-grained as well as microscopic models to directly break time-symmetry of the solution) is only included on the level of the initial conditions, while the solution of the macroscopic equations is still time-reversible. We have provided numerical evidence that random initial conditions lead to the emergence of pseudo time-irreversible solutions of the macroscopic equations. First, the hypothesis of spectral equipartition was verified, showing that the system irreversibly converges to a state where the kinetic energy is evenly spread across the Fourier modes of the momentum density (i.e. the relevant degrees of freedom) in time average, which is expected in thermodynamic equilibrium. Further numerical simulations indicated the presence of pseudo-irreversible transport of heat and momentum in systems supplemented with initial macroscopic order. Interestingly, though long-wavelength order was indeed found to be irreversibly decaying, the form of the relaxation had non-diffusional characteristics. There can be various reasons of the discrepancy, including (i) the lack of finite-scale corrections (potentially leading to the appearance of time-symmetry breaking terms), (ii) numerical reasons, or (iii) the incompleteness of the classical microscopic dynamics. Counter-arguments for the first two possible reasons can be that dissipative terms are also absent in the exact finite-scale equations of Olla, Varadhan and Yau \cite{Olla1993}, and that the macroscopic dynamical equations here are scale-free, and therefore no qualitatively different behaviour is expected to be observed for different system sizes and discretisation parameters. In contrast, the idea of incompleteness of the classical dynamical equations is supported by quantum mechanical arguments \cite{PhysRevA.42.78,PhysRevA.40.1165}, and also by the fact that imposing direct time-symmetry breaking on the level of the dynamical equations in the derivation of coarse-grained models in statistical physics also results in diffusional relaxation for mass, momentum, and energy on macroscopic scales.} These suggest that a Markovian stochastic component might be missing in the Hamiltonian particle dynamics \cite{WignerFunctions}, which is practically invisible in small systems, but naturally ensures true time irreversibility as well as the diffusive relaxation of mass, momentum and kinetic energy on macroscopic scales.

\section*{Acknowledgements}

The author wishes to thank M. te Vrugt (University of M\"unster, Germany), A. J. Archer and K. Khusnutdinova (Loughborough Univeristy, UK) for their valuable comments.

\appendix

\section{Microscopic dynamical equation}

\subsection{Definitions}

The analytical calculations refer to 3 spatial dimensions. Accordingly, the Fourier/inverse Fourier transforms and the Dirac-delta distribution are defined as: 

\begin{itemize}
\item Fourier transform: $\displaystyle{F(\mathbf{k})\equiv\frac{1}{(2\,\pi)^3} \int d\mathbf{r} \left\{ f(\mathbf{r}) e^{-\imath\,\mathbf{k}\cdot\mathbf{r}} \right\}}$;
\item Inverse Fourier transform: $\displaystyle{f(\mathbf{r})\equiv \int d\mathbf{k} \left\{ F(\mathbf{k}) e^{\imath\,\mathbf{k}\cdot\mathbf{r}} \right\}}$;

\item $3$-dimensional Dirac-delta: $\displaystyle{\delta(\mathbf{r}) \equiv \delta(x)\delta(y)\delta(z) = \frac{1}{(2\pi)^3} \int d\mathbf{k} \left\{ e^{\imath\,\mathbf{k}\cdot\mathbf{r}}\right\}}$ ;

\item The Fourier transform of $f(\mathbf{r})\equiv\delta(\mathbf{r}-\mathbf{a})$ reads: $\displaystyle{F(\mathbf{k}) = \frac{1}{(2\,\pi)^3}}\,e^{-\imath\,\mathbf{k}\cdot\mathbf{a}}$.

\end{itemize}

\subsection{Mass and momentum balance}

Let $m$ be the particle mass, and let $\mathbf{r}_i(t)$ and $\mathbf{p}_i(t)$ be the position and momentum of particle $i$, respectively. The microscopic mass and momentum densities read:
\begin{itemize}
\item Mass density: $\displaystyle{\hat{\rho}(\mathbf{r},t)= m \sum_i\delta[\mathbf{r}-\mathbf{r}_i(t)]}$;\\
Fourier transform: $\displaystyle{\hat{P}(\mathbf{k},t)= \frac{m}{(2\,\pi)^3} \sum_i e^{-\imath\,\mathbf{k}\cdot\mathbf{r}_i(t)}}$; 
\item Momentum density: $\displaystyle{\hat{\mathbf{g}}(\mathbf{r},t)=\sum_i \mathbf{p}_i(t)\,\delta[\mathbf{r}-\mathbf{r}_i(t)]}$;\\
Fourier transform: $\displaystyle{\hat{\mathbf{G}}(\mathbf{k},t)= \frac{1}{(2\,\pi)^3} \sum_i \mathbf{p}_i(t) \, e^{-\imath\,\mathbf{k}\cdot\mathbf{r}_i(t)}}$; 
\end{itemize}
In the derivation of the microscopic continuum equations, we use the canonical equations $\mathbf{p}_i(t)=m\,\dot{\mathbf{r}}_i(t)$ and $\dot{\mathbf{p}}_i(t)=-\partial V/\partial \mathbf{r}_i(t)$, where $V=\frac{1}{2}\sum_{i,j}u(|\mathbf{r}_i(t)-\mathbf{r}_j(t)|)$ is the potential energy, and $u(r)$ the isotropic pair potential. The time derivative of the Fourier transform of the microscopic mass density reads:
\begin{equation}
\frac{\partial \hat{P}(\mathbf{k},t)}{\partial t} = -\frac{\imath\,\mathbf{k}}{(2\,\pi)^3} \cdot \sum_i m\,\dot{\mathbf{r}}_i(t)\,e^{-\imath\,\mathbf{k}\cdot\mathbf{r}_i(t)} = -\imath\,\mathbf{k}\cdot\hat{\mathbf{G}}(\mathbf{k},t) \enskip ,
\end{equation}
thus indicating
\begin{equation}
\frac{\partial \hat{P}(\mathbf{k},t)}{\partial t} + \imath\,\mathbf{k}\cdot\hat{\mathbf{G}}(\mathbf{k},t) = 0 \quad \Rightarrow \quad  \frac{\partial \hat{\rho}(\mathbf{r},t)}{\partial t} + \nabla\cdot\hat{\mathbf{G}}(\mathbf{r},t) = 0 \enskip .
\end{equation}
The time derivative of the Fourier transform of the microscopic momentum density reads:
\begin{equation}
\frac{\partial \hat{\mathbf{G}}(\mathbf{k},t)}{\partial t} = \frac{1}{(2\,\pi)^3}\sum_i \left\{ \dot{\mathbf{p}}_i(t) - [\imath\,\mathbf{k}\cdot\dot{\mathbf{r}}_i(t)]\,\mathbf{p}_i(t) \right\} e^{-\imath\,\mathbf{k}\cdot\mathbf{r}_i(t)} \enskip ,
\end{equation}
thus yielding
\begin{equation}
\displaystyle{\frac{\partial \hat{\mathbf{G}}(\mathbf{k},t)}{\partial t} + \frac{\imath\,\mathbf{k}}{(2\,\pi)^3} \cdot \sum_i \frac{\mathbf{p}_i(t) \otimes \mathbf{p}_i(t)}{m}  e^{-\imath\,\mathbf{k}\cdot\mathbf{r}_i(t)}} = -\frac{1}{(2\,\pi)^3}\sum_i \frac{\partial V}{\partial \mathbf{r}_i(t)}e^{-\imath\,\mathbf{k}\cdot\mathbf{r}_i(t)} \enskip .
\end{equation}
Assuming that the Fourier transform of the pair potential [denoted by $U(k)$] exists, the pair potential can be written as:
$\displaystyle{ u(r) = \int d\mathbf{k}\,\left\{ U(k) e^{\imath\,\mathbf{k}\cdot\mathbf{r}} \right\} }$. Consequently,
\begin{equation}
\begin{split}
\frac{\partial V}{\partial \mathbf{r}_i(t)} & = \frac{\partial}{\partial\mathbf{r}_i(t)} \left(\frac{1}{2}\sum_{m,n} \int d\mathbf{q} \, \left\{ U(q)\, e^{\imath\,\mathbf{q}\cdot[\mathbf{r}_m(t)-\mathbf{r}_n(t)]} \right\} \right)\\
& = \frac{1}{2} \int d\mathbf{q}\,\left\{ U(q) \sum_{m,n} \imath\,\mathbf{q}\,(\delta_{m,i}-\delta_{n,i})\,e^{\imath\,\mathbf{q}\,\cdot[\mathbf{r}_m(t)-\mathbf{r}_n(t)]}  \right\} \\
&  = \int d\mathbf{q} \, \left\{ \imath\,\mathbf{q}\,U(q) \sum_j e^{\imath\,\mathbf{q}\cdot[\mathbf{r}_i(t)-\mathbf{r}_j(t)]} \right\} \enskip .
\end{split}
\end{equation}
Introducing $\displaystyle{\hat{\underline{\underline{K}}}(\mathbf{k},t) \equiv \frac{1}{(2\,\pi)^3} \cdot \sum_i \frac{\mathbf{p}_i(t) \otimes \mathbf{p}_i(t)}{m}  e^{-\imath\,\mathbf{k}\cdot\mathbf{r}_i(t)}}$, and using the above expression in the spectral momentum equation yield:
\begin{equation}
\label{App6}
\begin{split}
\frac{\partial \hat{\mathbf{G}}(\mathbf{k},t)}{\partial t} + \imath\,\mathbf{k}\cdot\hat{\underline{\underline{K}}}(\mathbf{k},t) &= -\frac{1}{(2\,\pi)^3}\int d\mathbf{q} \left\{ \imath\,\mathbf{q}\,U(q) \sum_{i,j} e^{-\imath\,(\mathbf{k}-\mathbf{q})\cdot\mathbf{r}_i(t)-\imath\,\mathbf{q}\cdot\mathbf{r}_j(t)} \right\} \\
& = -\frac{(2\,\pi)^3}{m^2} \int d\mathbf{q} \left\{ \imath\,\mathbf{q}\,U(q) \, \hat{P}(\mathbf{q},t) \, \hat{P}(\mathbf{k}-\mathbf{q},t)\right\} \enskip .
\end{split}
\end{equation}
Calculating the inverse Fourier transform of both sides yields:
\begin{equation}
\label{App7}
\frac{\partial \hat{\mathbf{g}}(\mathbf{r},t)}{\partial t} + \nabla\cdot \mathbb{K}(\mathbf{r},t) = -\hat{\rho}(\mathbf{r},t) \nabla \int d\mathbf{r}' \,\left\{ \frac{u(\mathbf{r}')}{m^2}\,\hat{\rho}(\mathbf{r}-\mathbf{r}',t) \right\} \enskip ,
\end{equation}
where $\displaystyle{ \hat{\mathbb{K}}(\mathbf{r},t)=\sum_i \frac{\mathbf{p}_i(t) \otimes \mathbf{p}_i(t)}{m}\delta[\mathbf{r}-\mathbf{r}_i(t)]  }$ is the microscopic kinetic stress tensor. 

\subsection{Global momentum conservation}

To show that Eq. (\ref{App7}) conserves momentum, we start from an equivalent formulation of Eq. (\ref{App6}):
\begin{equation}
\label{App8}
\frac{\partial \hat{\mathbf{G}}}{\partial t} + \imath\,\mathbf{k}\cdot\underline{\hat{\underline{K}}} = (2\,\pi)^3 [(\mathbf{F}\,\hat{P}) * \hat{P}] \enskip ,
\end{equation}
where $\mathbf{F}(\mathbf{k})=-\imath\,\mathbf{k}\,U(k)/m^2$, and $*$ stands for convolution in the wave number space. Using the convolution theorem, the right-hand side can be written as:
\begin{equation}
\begin{split}
[(\mathbf{F}\,\hat{P})&  * \hat{P}](\mathbf{k},t) =\\
&  \frac{1}{2}\int d\mathbf{q} \left\{  \mathbf{F}(\mathbf{q})\hat{P}(\mathbf{q},t)\hat{P}(\mathbf{k}-\mathbf{q},t) + \mathbf{F}(\mathbf{k}-\mathbf{q})\hat{P}(\mathbf{k}-\mathbf{q},t)\hat{P}(\mathbf{q},t) \right\} \enskip .
\end{split}
\end{equation}
Setting $\mathbf{k}\equiv\mathbf{0}$ in the equation yields:
\begin{equation}
[(\mathbf{F}\,\hat{P}) * \hat{P}](\mathbf{0},t) = \frac{1}{2}\int d\mathbf{q}\,\{ [\mathbf{F}(\mathbf{q})+\mathbf{F}(-\mathbf{q})] |\hat{P}(\mathbf{q},t)|^2\} = \mathbf{0} \enskip ,
\end{equation}
since $\mathbf{F}(\mathbf{q})+\mathbf{F}(-\mathbf{q}) = \mathbf{0}$ (Newton's third law). Setting $\mathbf{k} = \mathbf{0}$ in Eq. (\ref{App8}) then yields
\begin{equation}
\frac{d \hat{\mathbf{G}}(\mathbf{0},t)}{d t} = \mathbf{0} \enskip .
\end{equation}
Since the total momentum of the system is $\mathbf{P}(t) = (2\,\pi)^3\hat{\mathbf{G}}(\mathbf{0},t)$, the time derivative of the total momentum is zero: $d\mathbf{P}(t)/dt = 0$, i.e., the total momentum of the system is conserved.

\subsection{Microscopic origin of the local Gibbs-Duhem relation}

The potential energy of the system can be directly expressed in terms of the pair potential and the microscopic mass density as follows:
\begin{equation}
\begin{split}
V &= \frac{1}{2}\sum_{i,j} u(|\mathbf{r}_i-\mathbf{r}_j|) = \frac{1}{2} \int d\mathbf{k} \left\{ U(k) \sum_{i,j} e^{\imath\,\mathbf{k}\cdot[\mathbf{r}_i(t)-\mathbf{r}_j(t)]} \right\} \\
& = \frac{1}{2} \left[\frac{(2\pi)^3}{m}\right]^2 \int d\mathbf{k} \left\{ U(k) \hat{P}(\mathbf{k},t) \hat{P}(-\mathbf{k},t) \right\} = \\
& = \frac{1}{2} \int d\mathbf{r}' \int d\mathbf{r}''\int d\mathbf{r} \left\{ \frac{u(r')}{m^2} \hat{\rho}(\mathbf{r}'',t) \hat{\rho}(\mathbf{r},t) \left[\frac{1}{(2\pi)^3}\int d\mathbf{k} \left\{e^{\imath\,\mathbf{k}\cdot(\mathbf{r}'+\mathbf{r}''-\mathbf{r})} \right\} \right] \right\} =\\
& = \frac{1}{2} \int d\mathbf{r} \left\{ \hat{\rho}(\mathbf{r},t)  \int d\mathbf{r}' \left[ \frac{u(r')}{m^2} \hat{\rho}(\mathbf{r}-\mathbf{r}',t) \right] \right\} = \frac{1}{2} \int d\mathbf{r} \left\{ \hat{\rho} \, \left( \frac{u}{m^2}*\hat{\rho} \right) \right\} \enskip ,  
\end{split}
\end{equation}
where $*$ stands for spatial convolution. Since $u(r)$ is radially symmetric, $u(|\mathbf{r}-\mathbf{r}'|)=u(|\mathbf{r}'-\mathbf{r}|)$. Using this together with the convolution theorem results in the first functional derivative of $V$ with respect to $\hat{\rho}$:
\begin{equation}
\frac{\delta V}{\delta\hat{\rho}} = \frac{u}{m^2} * \hat{\rho} \enskip .
\end{equation}
Using this result in Eq. (\ref{App7}) gives:
\begin{equation}
\frac{\partial \hat{\mathbf{g}}}{\partial t} + \nabla\cdot\hat{\underline{\underline{K}}} = -\hat{\rho} \nabla \frac{\delta V}{\delta \hat{\rho}} \enskip .
\end{equation}

\subsection{Microscopic closure relation}

To find a closure relation for the microscopic kinetic stress, we show that $\hat{\mathbb{K}}(\mathbf{r},t)$ can be related to $\hat{\rho}(\mathbf{r},t)$ and $\hat{\mathbf{g}}(\mathbf{r},t)$:
\begin{equation}
\begin{split}
\hat{\rho}(\mathbf{r},t)\, \hat{\mathbb{K}}(\mathbf{r},t) & - \hat{\mathbf{g}}(\mathbf{r},t) \otimes \hat{\mathbf{g}}(\mathbf{r},t) \\ & = \sum_{i,j} [\mathbf{p}_j(t)\otimes \mathbf{p}_j(t) - \mathbf{p}_i(t)\otimes \mathbf{p}_j(t) ] \delta[\mathbf{r}-\mathbf{r}_i(t)]\delta[\mathbf{r}-\mathbf{r}_j(t)] \\
& = \frac{1}{2}\sum_{i,j}[\Delta \mathbf{p}_{ij}(t)\otimes \Delta\mathbf{p}_{ij}(t)]\,\delta[\mathbf{r}-\mathbf{r}_i(t)]\delta[\mathbf{r}-\mathbf{r}_j(t)] \\
& = \frac{1}{2}\sum_{i,j} [\Delta \mathbf{p}_{ij}(t)\otimes \Delta\mathbf{p}_{ij}(t)]\,\delta[\Delta\mathbf{r}_{ij}(t)]\delta[\mathbf{r}-\bar{\mathbf{r}}_{ij}(t)] \enskip ,
\end{split}
\end{equation}
where $\Delta \mathbf{p}_{ij}(t) = \mathbf{p}_i(t)-\mathbf{p}_j(t)$, $\Delta \mathbf{r}_{ij}(t) = \mathbf{r}_i(t)-\mathbf{r}_j(t)$, $\bar{\mathbf{r}}_{ij}(t) = [\mathbf{r}_i(t)+\mathbf{r}_j(t)]/2$, and we used the identity $\delta(x-a)\delta(x-b)=\delta(a-b)\delta[x-(a+b)/2]$. This identity can be proven by interpreting the product of Dirac-delta functions as $\delta(x-a)\delta(x-b)=\lim_{\sigma\to 0}G_\sigma(x-a)G_\sigma(x-b)$, where $G_\sigma(x)$ is any nascent function to the Dirac-delta, i.e., $\delta(x)\equiv\lim_{\sigma \to 0}G_\sigma(x)$. Choosing $G_\sigma(x)\equiv\frac{1}{\sigma\sqrt{2\pi}}e^{-\frac{x^2}{2\,\sigma^2}}$, for instance, and introducing $s_\sigma(x)\equiv G_\sigma(x-a)G_\sigma(x-b)$ results in
\begin{equation}
\begin{split}
S_\sigma(k) & = \frac{1}{2\pi} \left\{ \frac{1}{\sqrt{4 \pi \sigma^2}} e^{- \frac{(a-b)^2}{4\,\sigma^2}} \right\} e^{- \left[\imath \frac{a+b}{2}k + \frac{k^2}{4}\sigma^2 \right]} \\
& \Rightarrow \quad \lim_{\sigma\to 0}S_\sigma(k) = \frac{1}{2\pi} \delta(a-b) e^{-\imath\,\frac{a+b}{2}\,k},
\end{split}
\end{equation}
which is the Fourier transform of $\delta(a-b)\delta[x-(a+b)/2]$. Since integration (inverse Fourier transform) and limit commute, it follows that $\delta(x-a)\delta(x-b)=\delta(a-b)\delta[x-(a+b)/2]$. Introducing $\hat{\mathbb{R}}(\mathbf{r},t)\equiv\hat{\rho}(\mathbf{r},t)\, \hat{\mathbb{K}}(\mathbf{r},t) - \hat{\mathbf{g}}(\mathbf{r},t) \otimes \hat{\mathbf{g}}(\mathbf{r},t)$ then yields:
\begin{equation}
\hat{\underline{\underline{R}}}(\mathbf{k},t) = \frac{1}{(2\pi)^3} \, \frac{1}{2}\sum_{i,j} [\Delta \mathbf{p}_{ij}(t)\otimes \Delta\mathbf{p}_{ij}(t)]\,\delta[\mathbf{r}_i(t)-\mathbf{r}_j(t)]\,e^{-\imath\,\mathbf{k}\cdot\bar{\mathbf{r}}_{ij}(t)} \enskip .
\end{equation}
Since $\mathbf{r}_i(t)\neq\mathbf{r}_j(t)$ for any $i \neq j$ (i.e., the particles cannot occupy the same position coincidentally), $\Delta \mathbf{p}_{ij}(t) \equiv 0$ for $i=j$, and $a\,\delta(x) = 0$ everywhere for $a=0$ (in the Fourier sense), the above expression is identically zero, which results in the microscopic closure relation
\begin{equation}
\hat{\rho}(\mathbf{r},t)\, \hat{\mathbb{K}}(\mathbf{r},t) = \hat{\mathbf{g}}(\mathbf{r},t) \otimes \hat{\mathbf{g}}(\mathbf{r},t) \enskip .
\end{equation}

\subsection{Instantaneous temperature}

The dimensional instantaneous temperature of the system is defined as:
\begin{equation}
T(t) \equiv \frac{2\,K(t)}{3\,k_B N} \enskip ,
\end{equation}
where $K(t)= \sum_i \frac{|\mathbf{p}_i(t)|^2}{2\,m}$ is the instantaneous kinetic energy of the system. By introducing the microscopic kinetic energy density $\hat{e}(\mathbf{r},t)\equiv\sum_i \frac{|\mathbf{p}_i(t)|^2}{2\,m}\delta[\mathbf{r}-\mathbf{r}_i(t)]$, the temperature can be expressed as
\begin{equation}
T(t) = \frac{2}{3\,k_B N} \int_V d\mathbf{r}\,\hat{e}(\mathbf{r},t) = \frac{m}{3\,k_B \rho_0} \left[\frac{1}{V} \int d\mathbf{r} \,\textrm{Tr}\,\hat{\mathbb{K}}(\mathbf{r},t) \right] \enskip , 
\end{equation}
where we used that the kinetic energy density is related to the kinetic stress tensor via $\hat{e}(\mathbf{r},t)= \frac{1}{2}\,\textrm{Tr}\,\hat{\mathbb{K}}(\mathbf{r},t)$. Applying the non-dimensionalisation yields:
\begin{equation}
\tilde{T}(t) = \frac{1}{3} \langle \textrm{Tr}\,\hat{\mathbb{K}}(\mathbf{r},t) \rangle \enskip ,
\end{equation}
where $\langle.\rangle$ stands for spatial average, $T=\tilde{T}T_0$ (where $T_0=\epsilon/k_B$ is the temperature scale), while $\mathbf{r}$, $t$ and $\hat{\mathbb{K}}$ are also dimensionless.

\section{Scaling limit}

\subsection{Dynamical equations}

Here we consider a system (called $\kappa$-system henceforth) with particle mass $\kappa^3 m$, pair potential $\kappa^3 u(r/\kappa)$, and the number density $n_0/\kappa^3$, where $\kappa$ is positive real. Using the \textit{same} units of length, time and energy as in case of the original system results in the following dimensionless dynamical equations:
\begin{eqnarray}
\partial_{t}\rho_\kappa + \nabla \cdot \mathbf{g}_\kappa &=& 0 \\
\partial_{t}\mathbf{g}_\kappa + \nabla \cdot \mathbb{K}_\kappa &=& -\rho_\kappa \nabla (v_\kappa * \rho_\kappa) \enskip ,
\end{eqnarray}
where 
\begin{equation}
v_\kappa(r) = \tilde{u}(\bar{l}r/\kappa)/\kappa^3 \enskip .
\end{equation}
The closure relation reads:
\begin{equation}
\rho_\kappa(\mathbf{r},t)\mathbb{K}_\kappa(\mathbf{r},t)=\mathbf{g}_\kappa(\mathbf{r},t) \otimes \mathbf{g}_\kappa(\mathbf{r},t) \enskip .
\end{equation}
It is trivial to see that the $\kappa$-system is equivalent to the original system under space-time rescaling, i.e., $\rho_\kappa(\mathbf{r},t)=\hat{\rho}(\mathbf{r}/\kappa,t/\kappa)$, $\mathbf{g}_\kappa(\mathbf{r},t)=\hat{\mathbf{g}}(\mathbf{r}/\kappa,t/\kappa)$ and $\mathbb{K}_\kappa(\mathbf{r},t)=\hat{\mathbb{K}}(\mathbf{r}/\kappa,t/\kappa)$. In addition, the spatial averages of the densities are preserved:
\begin{equation}
\begin{split}
\langle A_\kappa(\mathbf{r},t) \rangle & =\frac{1}{V} \int_{V} d\mathbf{r}\,A_\kappa(\mathbf{r},t) = \frac{1}{V} \int_{V} d\mathbf{r}\,\hat{A}(\mathbf{r}/\kappa,t/\kappa) \\
& = \frac{1}{V/\kappa^3} \int_{V/\kappa^3} d\mathbf{r}\,\hat{A}(\mathbf{r},t/\kappa) \equiv \langle \hat{A}(\mathbf{r},t/\kappa) \rangle \enskip ,
\end{split}
\end{equation}
where $A=\rho,\mathbf{g}$ or $\mathbb{K}$. Consequently, the temperature is also preserved, since
\begin{equation}
\label{App31}
\tilde{T}(t) = \frac{1}{3} \, \textrm{Tr}\,\langle\hat{\mathbb{K}}(\mathbf{r},t) \rangle = \frac{1}{3} \langle\textrm{Tr} \, \mathbb{K}_\kappa(\mathbf{r},\kappa t) \rangle
\end{equation}
for arbitrary $\kappa$. The key point of this re-scaling is that the physical quantities (mass and momentum densities, and the temperature) remain finite as $\kappa$ converges to $0$. The equivalence transformation indicates that the $O(\xi)$-scale dynamics of the $\kappa$-system corresponds to the $O(\xi/\kappa)$-scale dynamics of the original system. Consequently, the $O(\infty)$ (macroscopic) scale dynamics of the original system can be studied by studying the $O(1)$-scale dynamics of the $\kappa$-system in the $\kappa \to 0$ limit. The scaling limit of the dynamical equations reads:
\begin{eqnarray}
\label{App32}\partial_{t}\rho + \nabla \cdot \mathbf{g} &=& 0 \\
\label{App33}\partial_{t}\mathbf{g} + \nabla \cdot \mathbb{K} &=& -\rho \nabla (u_0 * \rho) \enskip ,
\end{eqnarray} 
where $A(\mathbf{r},t)\equiv\lim_{\kappa \to 0}A_\kappa(\mathbf{r},t)$ (with $A=\rho,\mathbf{g}$ and $\mathbb{K}$), while
\begin{equation}
\label{App34}
u_0(r) = \lim_{\kappa \to 0} \left[ \frac{1}{\kappa^3}\,\tilde{u}\left( \frac{\bar{l}\,r}{\kappa} \right) \right] \enskip .
\end{equation}
The macroscopic closure relation reads:
\begin{equation}
\lim_{\kappa \to 0} \left[ \rho_\kappa(\mathbf{r},t)\,\mathbb{K}_\kappa(\mathbf{r},t) \right] = \lim_{\kappa \to 0} \left[\mathbf{g}_\kappa(\mathbf{r},t) \otimes \mathbf{g}_\kappa(\mathbf{r},t) \right] \enskip .
\end{equation}
If $\rho(\mathbf{r},t)=\lim_{\kappa \to 0}\rho_\kappa(\mathbf{r},t)$ and $\mathbf{g}(\mathbf{r},t)=\lim_{\kappa \to 0}\mathbf{g}_\kappa(\mathbf{r},t)$ are bounded functions, then the above equation indicates that $\mathbb{K}(\mathbf{r},t)=\lim_{\kappa \to 0}\mathbb{K}_\kappa(\mathbf{r},t)$ is also bounded, and
\begin{equation}
\rho(\mathbf{r},t)\,\mathbb{K}(\mathbf{r},t) = \mathbf{g}(\mathbf{r},t)\otimes \mathbf{g}(\mathbf{r},t) \enskip .
\end{equation}
If, in addition, $\rho(\mathbf{r},t)>0$, the kinetic stress can be explicitly expressed in terms of the mass ans momentum densities:
\begin{equation}
\label{App37}
\mathbb{K}(\mathbf{r},t) = \frac{\mathbf{g}(\mathbf{r},t)\otimes \mathbf{g}(\mathbf{r},t)}{\rho(\mathbf{r},t)} \enskip .
\end{equation}
Alternatively, it is possible to introduce the macroscopic velocity field as the ratio of the macroscopic momentum and density fields: $\mathbf{v}(\mathbf{r},t)\equiv\mathbf{g}(\mathbf{r},t)/\rho(\mathbf{r},t)$. Eqs. (\ref{App32}) and (\ref{App33}) can be re-formulated as:
\begin{eqnarray}
\partial_t \rho + \nabla\cdot(\rho\,\mathbf{v}) &=& 0 \enskip ;\\
\partial_t \mathbf{v} + \mathbf{v}\cdot\nabla\mathbf{v} &=& -\nabla(u_0 * \rho) \enskip .
\end{eqnarray}
Finally, using Eq. (\ref{App37}) in Eq. (\ref{App31}) results in:
\begin{equation}
\tilde{T} = \frac{1}{3}\langle \rho|\mathbf{v}|^2 \rangle \enskip .
\end{equation}

\subsection{Scaling limit of the pair potential}

The scaling limit of the pair potential is given by Eq. (\ref{App34}). Introducing $\\\alpha\equiv\kappa/\bar{l}$ (where $0<\bar{l}<\infty$) results in:
\begin{equation}
u_0(r) = \bar{n}\,\lim_{\alpha \to 0}\left[ \tilde{u}(r/\alpha)/\alpha^3 \right] \enskip .
\end{equation}
The limit is evaluated as follows. The Fourier transform of $u_0(r)$ can be written as
\begin{equation}
U_0(k)=\frac{\bar{n}}{(2\,\pi)^3} \lim_{\alpha \to 0} F_\alpha(k) \enskip ,
\end{equation}
where
\begin{equation}
F_{\alpha}(k) = \int_0^\infty dr\left\{ (4 \pi r^2) \left[ \frac{1}{\alpha^3} \tilde{u}\left(\frac{r}{\alpha}\right) \right]\frac{\sin(k\,r)}{k\,r} \right\} \enskip ,
\end{equation}
where we utilised that the pair potential has radial symmetry. If
\begin{equation}
\lim_{\alpha \to 0} F_\alpha(k)=a_0 \quad \text{(constant)} \enskip ,
\end{equation}
then
\begin{equation}
\label{App45}
u_0(r) = \bar{n}\,a_0\,\delta(\mathbf{r}) \enskip .
\end{equation}
We note that the notation $\delta(\mathbf{r})$ in Eq. (\ref{App45}) is consistent, since the three-dimensional Dirac-delta distribution is radially symmetric. This can be shown by considering the radial Dirac-delta in three dimensions as the limit of a radially symmetric nascent function:
\begin{equation}
\delta(r) \equiv \lim_{\sigma \to 0} g_\sigma(r) \enskip ,
\end{equation} 
where $g_\sigma(r)=\frac{1}{(\sigma\sqrt{2\,\pi})^3}e^{-\frac{r^2}{2\,\sigma^2}}$, for instance. Since $\lim_{\sigma\to 0}G_\sigma(k) = 1/(2\,\pi)^3$, which is the Fourier transform of $\delta(x)\delta(y)\delta(z)$, $\delta(\mathbf{r})\equiv \delta(r)$. The same result can be obtained by using radial Fourier transform in general, however, one must be careful and bare in mind that $\int_0^\infty (4\,\pi\,r^2\,dr) \delta(r) = 1$, and therefore the Fourier transform of $\delta(r)$ is $\Delta(k) = \frac{1}{(2\,\pi)^3} \lim_{r \to 0}\frac{\sin(k\,r)}{k\,r} = \frac{1}{(2\,\pi)^3}$. As an example for the scaling limit of the par potential, we consider the Yukawa potential: $\tilde{u}(r)=\exp(-\gamma\,r)/r$. For this potential,
\begin{equation}
a_0=\lim_{\alpha \to 0} F_\alpha(k) = \lim_{\alpha\to 0} \frac{4\pi}{(k\,\alpha)^2+\gamma^2} = \frac{4\,\pi}{\gamma^2}
\end{equation}
Similarly, for the Hartee-Fock B potential $\tilde{u}(r)=\exp(-\gamma\,r^2-\delta\,r)$ the limit reads:
\begin{equation}
\begin{split}
F_\alpha(k) & = \frac{(\pi/\delta) ^{3/2} }{{2\,k'}} e^{\frac{(\gamma -\imath k')^2}{4 \delta }} \\
& \times \left\{(k'-\imath \gamma ) e^{\frac{i \gamma  k'}{\delta }} \text{erfc}\left(\frac{\gamma +\imath k'}{2 \sqrt{\delta }}\right)-(\gamma -\imath k') \left[\text{erfi}\left(\frac{k'+\imath \gamma }{2 \sqrt{\delta }}\right)-\imath\right]\right\} \enskip , 
\end{split}
\end{equation}
where $k'=\alpha\,k$. This indicates
\begin{equation}
a_0 = \lim_{\alpha \to 0} F_\alpha(k) = \frac{\pi ^{3/2}}{2\,\delta ^{5/2}} e^{\frac{\gamma ^2}{4 \delta }} \left(\gamma ^2+2 \delta \right) \text{erfc}\left(\frac{\gamma }{2 \sqrt{\delta }}\right)-\frac{\pi  \gamma }{\delta ^2} \enskip ,
\end{equation}
which is also finite. More complex integrals can be evaluated using computer software (such as Mathematica). For any pair potential for which $u_0(r)=\bar{n}\,a_0\,\delta(r)$ holds, Eqs. (38) and (39) indicate
\begin{eqnarray}
\partial_t \rho + \nabla\cdot(\rho\,\mathbf{v}) &=& 0 \enskip ; \\
\partial_t \mathbf{v} + \mathbf{v}\cdot\nabla \mathbf{v} &=& - c_0^2 \nabla \rho \enskip ,
\end{eqnarray}
were $c_0=\sqrt{\bar{n}\,a_0}$ is the group speed. Eliminating $c_0$ by measuring time in $\tau=1/c_0$ units results in the universal macroscopic continuum equations:
\begin{equation}
\label{App51}
\partial_t \rho + \nabla\cdot(\rho\,\mathbf{v}) = 0 \quad \quad \text{and} \quad \quad \partial_t \mathbf{v} + \mathbf{v}\cdot\nabla \mathbf{v} + \nabla \rho = 0 \enskip ,
\end{equation}
where the spatial average of $\rho(\mathbf{r},t)$ is unity. Finally, using the re-scaling in Eq. (25) results in the instantaneous dimensionless temperature
\begin{equation}
\label{App52}\Theta(t)\equiv\frac{3\,\tilde{T}(t)}{\bar{n}\,a_0} = \langle \rho|\mathbf{v}|^2\rangle \enskip .
\end{equation}

\section{Boundedness of the densities in the hydrodynamic limit}

\rp{To demonstrate the boundedness of the macroscopic scaling limit of the mass density,} first we consider a one-dimensional system where $p_i=0$ and 
\begin{equation}
\label{App53}
x_i=\left\{ \begin{matrix} i/(1-\Delta) & \text{for} \quad i<0 \\  0 & \text{for} \quad i=0 \\ i/(1+\Delta) & \text{for} \quad i>0 \end{matrix} \right. \enskip .
\end{equation}
In this system, the average density is trivially $1\mp\Delta$ for $x<0$ and $x>0$, respectively. This result can be recovered by taking the macroscopic limit of the microscopic mass density. The microscopic mass density of the system reads $\hat{\rho}(x)=\sum_{i=-\infty}^\infty \delta(x-x_i)$. The average density of the system reads:
\begin{equation}
\bar{\rho} = \lim_{L \to \infty} \frac{1}{2\,L} \int_{-L}^{+L} dx\,\hat{\rho}(x) = 1 - \lim_{L \to \infty} \frac{\{L(1-\Delta\})+\{L(1+\Delta)\}-1}{2\,L} = 1 \enskip ,
\end{equation}
where $\{.\}$ stands for the fractional part (since the nominator in the limit is bounded, the limit is zero). The density difference is defined as
\begin{equation}
\label{App55}
\delta\hat{\rho}(x) = \delta\hat{\rho}-\bar{\rho} =\left[ \sum_{i=-\infty}^{+\infty} \delta(x-x_i) \right] - 1 \enskip ,
\end{equation}
which guarantees $\int_{-\infty}^{+\infty}dx\,\delta\hat{\rho}(x) \equiv 0$. Using Eq. (\ref{App53}) in Eq. (\ref{App55}), the Fourier transform of $\delta\rho_\kappa(x)\equiv\delta\hat{\rho}(x/\kappa)$ reads:
\begin{equation}
\label{App56}
\delta P_\kappa(k) = \frac{\kappa}{2\,\pi} \sum_{i=-\infty}^{+\infty} e^{-\imath\,k\,\kappa\,x_i} = \frac{\imath\,\kappa}{4\,\pi}\csc\left[\frac{\kappa\,k}{2(\Delta-1)}\right]\csc\left[\frac{\kappa\,k}{2(\Delta+1)}\right]\sin\left(\frac{k\,\Delta\,\kappa}{\Delta^2-1}\right)
\end{equation} 
for $k \neq 0$, while $\delta \hat{P}(0)=0$. From Eq. (\ref{App56}),
\begin{equation}
\delta P(k)=\lim_{\kappa \to 0} \delta P_\kappa(k) = \frac{\imath\,\Delta}{k\,\pi} \enskip ,
\end{equation}
and therefore $\delta\rho(x) = \lim_{\kappa \to 0} \delta\rho_\kappa(x) =  \Delta\, \textrm{sgn}(x)$. Finally, the macroscopic density reads:
\begin{equation}
\rho(x) = 1+\delta\rho(x) = \left\{ \begin{matrix} 1-\Delta & \text{for} \quad x<0 \\ 1 & \text{for} \quad x=0 \\ 1+\Delta & \text{for} \quad x>0 \end{matrix} \right. \enskip ,
\end{equation}
as expected. The key result here is that the macroscopic density $\rho(x)$, which is an exact limit of the microscopic one, is a bounded function. Without loss of generality, we consider the following from of the dimensionless positions and momenta of the particles, respectively (see Fig. 1):
\begin{eqnarray}
\label{App59}\mathbf{r}_{j(i)}(t)&\equiv&\mathbf{r}_i^0+\mathbf{u}(\kappa\,\mathbf{r}_i^0,\kappa\,t)/\kappa + \Delta \mathbf{r}_{j(i)}(t) \\
\label{App60}\mathbf{p}_{j(i)}(t)&\equiv&\mathbf{w}(\kappa\,\mathbf{r}_i^0,\kappa\,t) + \Delta \mathbf{p}_{j(i)}(t)
\end{eqnarray}
where $\mathbf{r}_i^0 \in \mathbb{Z}^d$ is a lattice site on a $d$-dimensional uniform grid with grid spacing $h=1$ (the whole grid is covered as $i$ runs for the natural numbers), 
 $\mathbf{u}(,.,)$ and $\mathbf{w}(,.,)$ are smooth, zero-average vector fields generating the slowly varying components of the density and the momentum fields for $\kappa \ll 1$, $j(i)$ is an instantaneous map between the   lattice sites and the particle positions (a particle is assigned to the closest grid point on the deformed grid $\mathbf{r}_i^0+\mathbf{u}(\kappa \mathbf{r}_i^0,\kappa\,t)/\kappa$), while $\Delta \mathbf{r}_{j(i)}(t)$ and $\Delta \mathbf{p}_{j(i)}(t)$ carry the microscopic details. The Fourier transform of $\delta\rho_\kappa(\mathbf{r},t)\equiv\delta\hat{\rho}(\mathbf{r}/\kappa,t/\kappa)$ - where $\delta\hat{\rho}(\mathbf{r},t)=-1+\sum_{i} \delta[\mathbf{r}-\mathbf{r}_{i}(t)]$ - reads:
\begin{equation}
\label{App61}
\begin{split}
\delta P_\kappa(\mathbf{k},t) & = \left(\frac{\kappa}{2\,\pi}\right)^3 \sum_{i} e^{-\imath\,\mathbf{k}\,\kappa\,\mathbf{r}_{i}(t/\kappa)} - \frac{1}{(2\,\pi)^3}\int d\mathbf{r}\,e^{-\imath\,\mathbf{k}\cdot\mathbf{r}} \\
& = \frac{1}{(2\,\pi)^3} \left\{  \sum_{\mathbf{n}}\kappa^3 \,  e^{-\imath\,\mathbf{k}\cdot\left[ \kappa\,\mathbf{n} + \mathbf{u}(\kappa\,\mathbf{n},t) + \kappa\,\Delta\mathbf{r}_{j(\mathbf{n})}(t/\kappa) \right]}  - \int d\mathbf{r}\,e^{-\imath\,\mathbf{k}\cdot\mathbf{r}} \right\} \enskip ,
\end{split}
\end{equation}
where $\mathbf{n}=(n_1,n_2,n_3) \in \mathbb{Z}^3$. Eq. (\ref{App61}) indicates
\begin{equation}
\label{App62}\lim_{\kappa\to 0} \delta P_\kappa(\mathbf{k},t) = \frac{1}{(2\,\pi)^3} \int d\mathbf{r} \left[ e^{-\imath\,\mathbf{k}\cdot \mathbf{u}(\mathbf{r},t)} -1 \right] e^{-\imath\,\mathbf{k}\cdot\mathbf{r}} \enskip ,
\end{equation}
where we used that $\Delta\mathbf{r}_{j(\mathbf{n})}(t)$ is finite. The key idea in Eq. (\ref{App62}) is that the summation for the particles is converted into the Fourier transform in the zero Knudsen number limit. In one spatial dimension the argument reads:
\begin{equation}
\label{App63} e^{-\imath\,\mathbf{k}\cdot \mathbf{u}(\mathbf{r},t)} -1  = \sum_{p=1}^\infty \frac{[-\imath\,k \,u(x,t)]^p}{p!} = (-\imath\,k)\,u(x,t)+\frac{1}{2}(-\imath\,k)^2 u^2(x,t) + \dots ,
 \end{equation}
thus indicating 
\begin{equation}
\delta\rho(x,t) = \sum_{p=1}^\infty \frac{(-1)^p}{p!}\partial_x^p [u^p(x,t)]
\end{equation}
We note here that Eq. (\ref{App63}) is different in multiple dimensions, where matrix algebraic identities must be used to expand $(\mathbf{k}\cdot\mathbf{u})^p$. For $p=2$, for instance, the general form is $(\imath\,\mathbf{k}\cdot\mathbf{u})^2 = (\imath\,\mathbf{k}) \cdot [(\imath\,\mathbf{k}) \cdot (\mathbf{u} \otimes \mathbf{u})]$, and therefore the corresponding real-space operation is $\nabla\cdot[\nabla\cdot(\mathbf{u}\otimes\mathbf{u})]$, and so on. The key point in calculating the macroscopic limit of the mass density using Eq. (\ref{App59}) is that the microscopic details $\Delta\mathbf{r}_{j}(t)$ cancel in the macroscopic density. This, however, does not apply to the macroscopic momentum density. Using Eq. (\ref{App60}) in $\mathbf{g}_\kappa(\mathbf{r},t)=\hat{\mathbf{g}}(\mathbf{r}/\kappa,t/\kappa)$, where $\hat{\mathbf{g}}(\mathbf{r},t)=\sum_i \mathbf{v}_i(t)\,\delta[\mathbf{r}-\mathbf{r}_i(t)]$ is the microscopic momentum, yields:
\begin{equation}
\mathbf{G}_\kappa(\mathbf{k},t) = \left(\frac{\kappa}{2\,\pi}\right)^3 \sum_i \left[\mathbf{w}(\kappa\,\mathbf{n},t)+\Delta\mathbf{v}_{j(\mathbf{n})}(t/\kappa) \right]\,e^{-\imath\,\mathbf{k}\cdot[\kappa\,\mathbf{n}+\mathbf{u}(\kappa\,\mathbf{n},t)+\kappa\,\Delta\mathbf{r}_{j(\mathbf{n})}(t/\kappa)]} \enskip .
\end{equation}
Taking the $\kappa \to 0$ limit results in:
\begin{equation}
\mathbf{G}_\kappa(\mathbf{k},t) = \frac{1}{(2\,\pi)^3} \int d\mathbf{r}\, \left\{\left[\mathbf{w}(\mathbf{r},t)+\vec{\xi}(\mathbf{r},t)\right]\,e^{-\imath\,\mathbf{k}\cdot\mathbf{u}(\mathbf{r},t)}\right\}\,e^{-\imath\,\mathbf{k}\cdot\mathbf{r}} \enskip ,
\end{equation}
where $\vec{\xi}(\mathbf{r},t)=\lim_{\kappa to 0} \Delta\mathbf{v}_{j(\lfloor \mathbf{r}/\kappa \rfloor)}(t/\kappa)$. Considering only one spatial dimension and using Eq. (\ref{App59}), the macroscopic momentum density reads:
\begin{equation}
G(x,t) = \sum_{p=0}^\infty \frac{(-1)^p}{p!} \partial_x^p\{[w(x,t)+\xi(x,t)]u^p(x,t)\} \enskip .
\end{equation}

\bibliography{papers}

\end{document}